\documentclass[a4paper,11pt]{article}

\usepackage{latexsym}
\usepackage{amsfonts} 
\usepackage[mathscr]{euscript}
\usepackage{amsmath,mathrsfs,bm,amssymb,color, ascmac}

\title{\textsf{
Correlation inequalities for  Schr\"odinger operators 
}}

\date{\empty}
\author{
Tadahiro Miyao\\ 
 {\it Department of Mathematics,}
{\it Hokkaido University,}\\
{\it Sapporo 060-0810, Japan}\\
E-mail:
 miyao@math.sci.hokudai.ac.jp
}

\newcommand{\one}{{\mathchoice {\rm 1\mskip-4mu l} {\rm 1\mskip-4mu l}
{\rm 1\mskip-4.5mu l} {\rm 1\mskip-5mu l}}}
\newcommand{\h}{\mathfrak{H}}

\newcommand{\ex}{e}
\newcommand{\D}{\mathrm{dom}}

\newcommand{\la}{\langle}
\newcommand{\ra}{\rangle}
\newcommand{\Tr}{\mathrm{Tr}}

\newcommand{\BbbR}{\mathbb{R}}
\newcommand{\BbbN}{\mathbb{N}}
\newcommand{\BbbZ}{\mathbb{Z}}

\newcommand{\vepsilon}{\varepsilon}
\newcommand{\vphi}{\varphi}

\newcommand{\Cone}{\mathfrak{P}}

\newcommand{\no}{\nonumber \\}

\newcommand{\Bs}{{\boldsymbol s}}
\newcommand{\Bt}{{\boldsymbol t}}
\newcommand{\tomega}{\tilde{\omega}}

\begin{document}

\newtheorem{define}{Definition}[section]
\newtheorem{Thm}[define]{Theorem}
\newtheorem{Prop}[define]{Proposition}
\newtheorem{lemm}[define]{Lemma}
\newtheorem{rem}[define]{Remark}
\newtheorem{assum}{Condition}
\newtheorem{example}{Example}
\newtheorem{coro}[define]{Corollary}

\maketitle
\begin{abstract}
The purpose of the present paper is to 
analyze 
correlation structures of the ground states of the 
  Schr\"odinger operator.
We construct Griffiths inequalities for the ground state expectations by applying 
operator-theoretic correlation inequalities.
As an example of such an  application, we study the ground state properties of Schr\"odinger operators.

\begin{flushleft}
{\bf Mathematics Subject Classification (2010).} 
\end{flushleft}
Primary: 47D08, 82B10;
Secondary: 47N50, 47A63
\begin{flushleft}
{\bf
Keywords. 
} 
\end{flushleft}
Schr\"odinger operator; Correlation, Ground state expectation; Griffiths inequalities; Operator inequalities;
Self-dual cone.
\end{abstract} 
\section{Introduction}
\setcounter{equation}{0}

The so-called \lq\lq{}Ising model\rq\rq{} was introduced by Lenz \cite{Lenz} to study
 ferromagnetic properties of a magnet.
 This model was discussed in his PhD thesis by Ising \cite{Ising},  and has been actively studied by both
 mathematicians and physicists.
The Ising model on $\Lambda=[-L, L)^d\cap \BbbZ^d$ is defind as follows.
For each spin configuration ${\boldsymbol \sigma}=\{\sigma_x\}_{x\in
\Lambda} \in \Omega=\{-1, +1\}^{\Lambda}
$ on $\Lambda$, the energy of the Ising system is 
\begin{align}
H({\boldsymbol \sigma})=-\sum_{x, y\in \Lambda} J_{xy} \sigma_x\sigma_y,
\end{align} 
where $J_{xy}$ is a nonnegative coupling constant.
The thermal average is defined by 
\begin{align}
\la \sigma_A\ra=\sum_{{\boldsymbol \sigma\in \Omega}}\sigma_A e^{-\beta
 H({\boldsymbol \sigma})}\Big/ Z_{\beta},\ \ \
 Z_{\beta}=\sum_{{\boldsymbol \sigma} \in \Omega} e^{-\beta
 H({\boldsymbol \sigma})},
\end{align} 
where $\sigma_A=\prod_{x\in A} \sigma_x$ for each $A\subseteq \Lambda$.
In his study of  Ising ferromagnets \cite{Griffiths1, Griffiths2,
Griffiths3}, Griffiths discovered the well-known {\it Griffiths inequalities}.
 Kelly and Sherman refined  the Griffiths inequalities as follows \cite{KS}:
\begin{itemize}
\item First inequality: 
\begin{align}
\la \sigma_A\ra\ge 0,\ \ \ \ A\subseteq \Lambda;\label{First}
\end{align} 
\item Second inequality: \begin{align}
\la \sigma_A\sigma_B\ra-\la \sigma_A\ra\la \sigma_B\ra\ge 0,\ \ \ \ 
A, B\subseteq \Lambda.\label{Second}
\end{align} 
\end{itemize} 

 These inequalities played an important role
in the rigorous study of the Ising model \cite{Griffiths4}.
Accordingly, we can expect that the   Griffiths inequalities express the
essential idea of 
 correlation in the Ising system. 
Therefore, it is logical to ask  whether  similar inequalities hold 
for other  models. An attempt to  find a solution of  this question
can be regarded as an exploration of  the model-independent structure of correlations.
Ginibre's work \cite{Ginibre} was a first  important step toward understanding this model-independent 
structure.  His framework of constructing  the Griffiths inequalities 
still hold for several classical models  \cite{Percus}.   However, we  know of  a few examples of  quantum models
 that satisfy  Griffiths inequalities; it has been actively studied to construct the inequalities for quantum models, see, e.g., \cite{BBU, Kitatani,MNC}.

In recent studies, Miyao established the  Griffiths inequalities for both Bose and Fermi 
systems \cite{Miyao6}. His theory was constructed 
from the veiwpoint of operator-theoretic correlation inequalities. 
According to this theory, we can unify the method of  reflection positivity 
in the theory of phase transitions \cite{DLS, FILS, GJS},  Lieb's spin reflection positivity 
in the Hubbard model \cite{Lieb, Miyao2, Miyao7, Shen, Tian}
 and Griffiths inequalities. In this way, the new theory is expected to 
 describe a universal aspect of the  notion of correlation. 

 The Schr\"odinger operator is  
undoubtedly
 one of the most important models in quantum theory. 
Hence, we can expect that this model will provide a crucial clue, leading 
to better understanding of the  universal aspects of 
correlation.
 Conversely, there has been   little  research on  correlation structures of the ground states of 
 this model.\footnote{
For example, see \cite{AHS, LiebSimon}. In \cite{AHS}, Hydrogen-like
atoms in constant magnetic field are studied. In \cite{LiebSimon}, the
Born-Oppenheimer energy is investigated.
}
The principal aim of  the present paper is to analyze 
correlation properties of the ground states of 
the Schr\"odinger
operator in terms of  
the operator-theoretic correlation inequalities.
This kind of the study is expected to be useful, when we examine the entanglment structures of many-body systems.
Through this analysis, we clarify the  Griffiths inequalities for ground state  expectations.
 As we will see, our correlation inequalities provide qualitative information on the shape of ground states.
The forms of the  obtained results are consistent  with  (\ref{First}) and (\ref{Second}), as we will see in  Section \ref{Sec2}. This is more than coincidence because our construction is based on our previous work \cite{Miyao6} which is a generalization of the Griffiths inequalities. Finally,
remark  that our method can be applied  to 
many-body Schr\"odinger operators.

The remainder  of this paper is as follows.
In Section \ref{Sec2}, we display results from  the analysis of operator theoretic correlation
inequalities.

 In Section \ref{3}, we introduce several operator inequalities associated with self-dual cones.
As we will see, these inequalities are very useful to study 
correlation structures of the ground states.

Sections \ref{4}-\ref{8} are devoted to the analysis of the ground states of   Schr\"odinger
operators in terms of  
the theory constructed in Section \ref{3}.
 
In Appendix  \ref{AppA}, we construct a general theory of correlation inequalities as operator 
inequalities associated  with self-dual cones.
Although many of the results in this section  are already proved in 
previous studies \cite{Faris, Gross, Miyao1, Miyao2, Miyao4,
Miyao5, Miyao6, Miyao7, Miyao8}, we have specified them here   the  for readers' convenience.
\begin{flushleft}
{\bf Acknowledgments.}
The author is grateful to the anonymous referee for useful comments.
This work was partially supported by KAKENHI (20554421),   KAKENHI(16H03942)
and KAKENHI (18K0331508).
\end{flushleft}

\section{Results}\label{Sec2}
\setcounter{equation}{0}

\subsection{Definitions and assumptions}
We will study the Schr\"odinger operator
\begin{align}
H&=-\Delta_x- V
\end{align} 
acting on the Hilbert space $L^2(\BbbR^d; dx)$. 
As usual, $\Delta_x$  is a self-adjoint realization of the $d$-dimensional Laplacian, and $V$ is a potential. 

To state our results, we need the  assumptions {\bf (A)}, {\bf
(B)},  and {\bf (C)} below.

Our first assumption concerns the self-adjointness of $H$.

\begin{flushleft}

{\bf (A)} The potential $V:\BbbR^d \to \BbbR$ is  chosen such that 
$H$ is self-adjoint on $\D(-\Delta_x)$ and bounded from below. $\diamondsuit$
\end{flushleft} 

\begin{example}
{\rm
If $V\in L^n(\BbbR^d; dx)+L^{\infty}(\BbbR^d; dx)$ with $n=2$ for $d\le
 3$,
$n>2$ for $d=4$ and 
 $n=d/2$
 for $d\ge 4$, 
 then $V$ satisfies {\bf (A)},
see, e.g., \cite[Theorem X. 29]{ReSi2}. $\diamondsuit$
}
\end{example}

 As usual,  the Fourier transform of $f$ is defined  by 
\begin{align}
\hat{f}(p):=(2\pi)^{-d/2} \int_{\BbbR^d} dx\, e^{-i p\cdot x}
 f(x).
\end{align} 
Our second assumption is stated as  follows.

\begin{flushleft}
{\bf (B)} There exists an approximating sequence ${V_n}\neq 0$ for $V$ such
that
 (i)--(iii) hold: 
\end{flushleft}
\begin{itemize}
 \item[(i)] Let $H_n=-\Delta_x-V_n$. $H_n$
 converges to $H$ in the
strong resolvent sense as $n\to \infty$.\footnote{
Let $\{A_n\}_{n=1}^{\infty}$ be a sequence of self-adjoint operators on
	    $L^2(\BbbR^d;  dx)$.
We say that $A_n$  converges to $A$ in the  {\it strong resolvent sense} if $
(A_n-z)^{-1}
$ converges to $(A-z)^{-1}$ in the strong operator topology for all $z$ with $\mathrm{Im}
	    z \neq 0$.
}

\item[(ii)] 
For all $n\in \BbbN$ and a.e. $p$, the Fourier transform $\hat{V}_n(p)$
 exists and satisfies
$\hat{V}_n\in L^1(\BbbR^d; dp)$, $\hat{V}_n(p)
\ge 0 $ and $\hat{V}_n(-p)=\hat{V}_n(p)$ a.e. $p$.
Moreover, there exists an $\vepsilon>0$ such that $\mathrm{supp}\hat{V}_n \supset
	    B_{\vepsilon}(0)$, where $\mathrm{supp} \hat{V}_n=
\overline{\{p\in
	    \BbbR^d\, |\, \hat{V}_n(p)\neq 0\}}$ and
	    $B_{\vepsilon}(0)=\{p\in \BbbR^d\, |\, |p|<\vepsilon\}$.

 \item[(iii)] $\hat{V}_n(p)$ is monotonically increasing in $n$, i.e., $\hat{V}_n(p) \le
\hat{V}_{n+1}(p)$ a.e. $p$ for all $n\in \BbbN$. $\diamondsuit$
\end{itemize} 
\begin{rem}
{\rm 
In concrete applications, it often happens that $\hat{V}$ does not 
exist, or that   $\hat{V}$ exists,  but $\hat{V}\notin L^1(\BbbR^d;  dp)$.
Even in these cases, we can apply our theory of operator-theoretic
      correlation inequalities on  the basis of  the assumption
      {\bf (B)}.  This is the principal   reason for  introducing
 $\{V_n\}_{n=1}^{\infty}$. $\diamondsuit$
}
\end{rem} 

\begin{example}\label{Yukawa}
{\rm 
Let us consider  the  Yukawa potential,  $\displaystyle 
V(x)=\frac{e^{-m |x|}}{|x|}$ with $m> 0$.
In the three-dimensional case, we have
$\displaystyle 
\hat{V}(p)= \frac{ 2^{1/2}}{p^2+m^2}
$. Clearly, $\hat{V}(p)\notin L^1(\BbbR^3; dp)$.
In this case, we set
\begin{align}
V_n(x)=(2\pi)^{-3/2} \int_{\BbbR^3} e^{ip\cdot x} \hat{V}_n(p)dp,
\end{align} 
where 
\begin{align}
\hat{V}_n(p)=
\begin{cases}
\displaystyle 
\ \ \ \hat{V}(p) & \mbox{if $|p| \le n$}\\
\ \ \ 0 & \mbox{if $|p|>n$}
\end{cases}.
\end{align} 
Then,  $V_n$ satisfies the  assumption {\bf (B)}.\footnote{{\it Proof.} (ii) and (iii) of the condition {\bf (B)} are easy to check.

(i) Remark that $\|V_n-V\|_{L^2}=\|
\hat{V}_n-\hat{V}
\|_{L^2}\to 0$ as $n\to \infty$.
Thus, for each $\vphi\in C_0^{\infty}(\BbbR^3)$, we see that $
\|(V_n-V) \vphi\|_{L^2}\le \|V_n-V\|_{L^2} \|\vphi\|_{L^{\infty}} \to 0
$ as $n\to \infty$, which implies that $(-\Delta_x-V_n) \vphi$ 
 converges to $(-\Delta_x-V) \vphi$ as $n\to \infty$.
Because $C_0^{\infty}(\BbbR^3)$ is a common core for  $-\Delta_x-V_n$ and $-\Delta_x-V$,
we can apply a general theorem \cite[Theorem VIII. 25 (a)]{ReSi1}  and conclude that 
 $-\Delta_x-V_n$ converges to  $-\Delta_x-V$ in the strong resolvent sense. $\Box$}
 
 We can also deal with the case where $m=0$ by extending the above arguments.
 In this case, we have $\hat{V}(p)=2^{1/2}/p^2$. 
 Set $\hat{V}_n(p)=\hat{V}(p)\chi_{I_n}(p)$, where $\chi_{I_n}$ is the  indicator function of  a set  $I_n=B_n(0) \backslash B_{1/n}(0)$. Then we can readily confirm that $\hat{V}_n$ satisfies the assumption {\bf (B)}.
$\diamondsuit$
}
\end{example} 

\begin{example}\label{SchwartzEx}
{\rm 
We consider the three-dimensional case: $d=3$.
Let $V$ be a potential such that $V\in \mathcal{S}(\BbbR^3)$, the Schwartz class, and $V(-x)=V(x)$.
We assume that $\hat{V}(p) \ge 0$ for all $p\in \BbbR^3$.
Then the following properties are readily obtained:
\begin{itemize}
\item $\hat{V}(0)>0$;
\item $\hat{V}(p)$ is continuous in $p$.
\end{itemize}
Therefore, there exists a number $\vepsilon>0$ such that $\mathrm{supp} \hat{V} \supset B_{\vepsilon}(0)$. By setting $\hat{V}_n(p)=\hat{V}(p)$ for all $n\in \BbbN$, we see that the assumption {\bf (B)}
is satisfied.
A typical example is $V(x)=V_0 e^{-x^2/a^2}$ with $V_0>0$ and $a>0$.

}
\end{example}

For a given linear operator $A$, 
we denote by  $\sigma(A)$    spectrum of  $A$. 
The following assumption concerns the least eigenvalue of $H$.
\begin{flushleft}
{\bf (C)}
There exists an $n_0\in \BbbN$ such that, for all $n\ge n_0$,
$\inf \sigma(H_n)$ is an eigenvalue of $H_n$.
 In addition,  $\inf \sigma(H)$ is an eigenvalue of $H$. $\diamondsuit$

\end{flushleft}

\begin{example}
{\rm 
Let us consider the Yukawa potential given in Example \ref{Yukawa}.
If $m$ is small, then $\inf \sigma(H)$ is an eigenvalue. This is  because $H^{(m)}$ converges to $H^{(m=0)}$
,
the Hamiltonian of the hydrogen-like atom, as $m\to +0$ in the strong resolvent sense.
Here, we clarify the $m$-dependence of $H$ by expressing $H$ as $H^{(m)}$.
Since $H_n$ converges to $H$ in the strong resolvent sense as $n\to \infty$, 
$\inf \sigma(H_n)$ must be an eigenvalue, provided that $n$ is large enough. $\diamondsuit$
}
\end{example}

\begin{define}
{\rm 
We say that the potential $V$ is {\it regular} if it satisfies  
 {\bf (A)}, {\bf (B)},  and {\bf (C)}. $\diamondsuit$
}
\end{define} 

\begin{example}
{\rm 
\begin{itemize}
\item[(i)] The Yukawa potential discussed in Example \ref{Yukawa} is regular, if $m=0$ or $m$ is sufficiently small.
\item[(ii)]
Let us consider the potantial $V$ concretely given in Example \ref{SchwartzEx}.
Then $V$ is regular provided that $V_0$ is large enough. $\diamondsuit$
\end{itemize}
 }
\end{example}

\begin{define}
{\rm 
Let $A$ be a self-adjoint operator,  bounded from below.
If $\inf \sigma(A)$ is an eigenvalue, then  the corresponding 
normalized eigenvectors are 
called {\it ground states} of $A$. $\diamondsuit$
}
\end{define}

The following proposition is a basic input.

\begin{Prop}\label{GSUni}
Assume that $V$ is regular. 
The ground state of $H$ (resp.,  $H_n$) is unique. Let $\psi$
 (resp.,  $\psi_n$)
 be the unique ground  state of $H$ (resp.,  $H_n$).  We have the following:
\begin{itemize}
\item[{\rm (i)}] $\psi(x)>0$ and $\psi_n(x)>0$ a.e. $x$.
\item[{\rm (ii)}] $\hat{\psi}(p)>0$ and $\hat{\psi}_n(p)>0$ a.e. $p$.
\end{itemize} 
\end{Prop} 

\begin{rem}
{\rm 
The property (i) is well-known, see, e.g., \cite[Theorem XIII.45]{ReSi4},
 however, as far as we know, the property (ii) is novel. $\diamondsuit$
}
\end{rem} 

We  prove Proposition \ref{GSUni} in Section \ref{4}.

We denote by  $
\mathscr{B}(\mathfrak{H})$ the set of all bounded linear operators on a Hilbert space  $\mathfrak{H}$.
\begin{define}
{\rm 
Let $\psi$ (resp.,  $\psi_n$) be the unique ground state of $H$
 (resp.,  $H_n$).
For each $A\in \mathscr{B}(L^2(\BbbR^d; dx))$, we define the 
{\it ground
 state expectation} $\la A\ra$ by 
\begin{align}
\la A\ra=\la \psi|A\psi\ra.
\end{align} 
Similarly, we define $\la A\ra_n=\la\psi_n|A\psi_n\ra$. $\diamondsuit$
}
\end{define}

\subsection{First inequalities}

 In this study, we   write the operator  $M_f$,  for  
 multiplication  by  the function $f$, simply  as $f$, 
 if no confusion occurs.

For each $f\in L^{\infty}(\BbbR^d; dx)$, a linear operator $f(-i
 \nabla_x)$
 is defined by 
\begin{align}
f(-i \nabla_x) \phi=\Big(
f(p) \hat{\phi}
\Big)^{\vee},\ \ \ \ \phi\in L^2(\BbbR^d; dx),\label{DefNabla}
\end{align} 
where $\vee$ is the inverse Fourier transform.
 
Let
\begin{align}
\mathfrak{A}&=\big\{
f\in L^{\infty}(\BbbR^d; dx)\cap L^2(\BbbR^d; dx)\, \big|\, \hat{f}(p)\ge 0\ \
 \mbox{a.e. $p$}
\big\}.
\end{align}
The following theorem corresponds to the first Griffiths inequality (\ref{First}).

\begin{Thm}\label{FirstInq}
Assume that $V$ is regular.
\begin{itemize}
\item[{\rm (i)}]
For all $f\in \mathfrak{A}$,  $\la f \ra\ge 0$.
The equality holds if and only if $f=0$.
\item[{\rm (ii)}]
For all $f\in \mathfrak{A}$,   $\big\la f(-i \nabla_x) \big\ra\ge 0$.
The equality holds if and only if $f=0$.  
\end{itemize} 
\end{Thm}  

We prove Theorem \ref{FirstInq} in Section  \ref{4}.

\subsection{Second inequalities}

Here, we state some results related to  the second Griffiths inequality
(\ref{Second}).
For this purpose, we introduce the following:
\begin{align}
\mathfrak{A}_{\mathrm{e}}&=\big\{
f\in L^{\infty}(\BbbR^d;dx)\cap L^2(\BbbR^d; dx)\, \big|\, \hat{f}(p)\ge 0\ \
 \mbox{a.e. $p$ and $f(-x)=f(x)$ a.e. $x$}
\big\}.
\end{align}

By the assumption (i) of {\bf (B)}, we can readily expect that $\la A\ra_n$ converges to $\la A\ra$ as $n\to \infty$.
The following theorem provides more detailed information on the convergence.

\begin{Thm}\label{MonoEx}
Assume that $V$ is regular. 
\begin{itemize}
\item[{\rm (i)}]For all $f\in \mathfrak{A}_{\mathrm{e}}$, $\la f\ra_n$
is monotonically increasing in $n$  and converges to $\la f\ra$.
\item[{\rm (ii)}]For all $f\in \mathfrak{A}_{\mathrm{e}}$,  
	     $\big\la f(-i \nabla_x)\big\ra_n$ is monotonically decreasing in
	     $n$ and converges to $\big\la f(-i\nabla_x)\big\ra$.
\end{itemize} 
\end{Thm} 

We  provide a proof  of Theorem \ref{MonoEx} in Section  \ref{5}.

The following theorem is a generalization of the second Griffiths inequality (\ref{Second}).
\begin{Thm}\label{GriSec}
Assume that $V$ is regular. For all $f, g\in \mathfrak{A}_{\mathrm{e}}$,
we have the following:
\begin{itemize}
\item[{\rm (i)}] $\la fg\ra-\la
 f\ra \la g\ra\ge 0$.
\item[{\rm (ii)}]
	     $\big\la f(-i \nabla_x)g(-i \nabla_x)\big\ra
-\big\la
 f(-i \nabla_x)\big\ra 
\big\la g(-i \nabla_x)\big\ra\ge 0$.
\item[{\rm (iii)}]
$
\big\la f(-i \nabla_x) g\big\ra
-\big\la
 f(-i \nabla_x)\big\ra 
\big\la g \big\ra\le 0
$.
\end{itemize} 
\end{Thm} 
\begin{rem}\label{RealPRem}
{\rm 
In Section \ref{PfRem}, we will show the following:
\begin{itemize}
\item[(i)] $
\big\la f(-i \nabla_x) g\big\ra
$ is a real number;
\item[(ii)]  if $f(x) \ge 0$ or $g(x) \ge 0$, then $
\big\la f(-i \nabla_x) g\big\ra \ge 0
$. 
\end{itemize}
Thus, Theorem \ref{GriSec} (iii) is meanigful. $\diamondsuit$
}
\end{rem} 

We provide  a proof of Theorem \ref{GriSec} in Section \ref{6}.

\begin{define}{\rm 
Let $V^{(1)}$ and $ V^{(2)}$ be regular potentials. Let $\hat{V}_n^{(1)}$ and
 $\hat{V}^{(2)}_n$
be the  corresponding approximating  functions appearing in   the condition {\bf (B)}.
We write $V^{(1)} \succeq V^{(2)}$, if
 there
exists an $n_0\in \BbbN$ such that for all $n\ge n_0$, 
 $\hat{V}_n^{(1)}(p) \ge \hat{V}_n^{(2)}(p)$ a.e. $p$. $\diamondsuit$
}
\end{define} 

\begin{example}\label{Potential1}
{\rm 
Let $W$ be a regular potential. Assume that $\lambda W$ is regular for
 all $\lambda\in I$, where $I$ is an open subset of $(0, \infty)$.
We set $V^{(1)}=\lambda_1 W$ and $V^{(2)}=\lambda_2 W$. 
If $\lambda_1, \lambda_2\in I$ and $\lambda_1\ge \lambda_2$,
then $V^{(1)} \succeq V^{(2)}$.  

As a  typical example, we consider the following. Let $W$ be a potential given in Example \ref{SchwartzEx}:
$W(x)=W_0e^{-x^2/a^2}$. The potential $W$ is regular provided that $W_0$ is large enough.
Let $I$ be an open subset of $\BbbR$ such that $I\subseteq [1, \infty)$. Then $\lambda W$
 is regular for all $\lambda\in I$.  
$\diamondsuit$
}
\end{example}

Let $V^{(1)}$ and $V^{(2)}$ be regular potentials. We consider Schr\"odinger
operators given by
\begin{align}
H^{(1)}=-\Delta_x-V^{(1)},\ \ \ \ H^{(2)}=-\Delta_x-V^{(2)}.
\end{align}  
Let $\psi^{(1)}$ (resp.,  $\psi^{(2)}$) be the unique ground state of $H^{(1)}$
(resp.,  $H^{(2)}$). We set
\begin{align}
\la A \ra^{(1)}=\big\la \psi^{(1)} |A\psi^{(1)}\big\ra,\ \ \  \la A\ra^{(2)}=\big\la \psi^{(2)}|A\psi^{(2)}\big\ra.
\end{align}

In Section \ref{7}, we demonstrate the following.
\begin{Thm}\label{MonoG}
Assume that $V^{(1)}$ and $ V^{(2)}$ are  regular.
\begin{itemize}
\item[{\rm (i)}]If $V^{(1)} \succeq V^{(2)}$, then 
 $\la f\ra^{(1)}\ge \la f\ra^{(2)}$ for all $f\in
 \mathfrak{A}_{\mathrm{e}}$.

\item[{\rm (ii)}]
If $V^{(1)} \succeq V^{(2)}$, then 
 $\big\la f(-i \nabla_x)\big\ra^{(1)}\le \big\la f(-i \nabla_x)\big\ra^{(2)}$ for all $f\in
 \mathfrak{A}_{\mathrm{e}}$.
\end{itemize} 
\end{Thm} 
\subsection{Application I: Ground state properties}\label{SecApp1}
We study some properties of the ground states by the correlation inequalities.
In Section \ref{NewApp}, we will show the following theorems.

\begin{Thm}\label{PPGS1}
Assume that $V$ is regular. Let $\psi$ be the ground state of $H$.
We set 
\begin{align}
\mathcal{C}(V)&=\{x\in \BbbR^d\, |\, \mbox{$\psi$ is continuous  at $x$}\},\\
\hat{\mathcal{C}}(V)&=\{p\in \BbbR^d\, |\, \mbox{$\hat{\psi}$ is continuous  at $p$}\}.
 \end{align} 
Assume that   
$0\in   \mathcal{C}(V)$ and $0\in   \hat{\mathcal{C}}(V)$.
 Then we have the following:
\begin{itemize}
\item[{\rm (i)}] $\psi(0)\ge \psi(x)$  for all  $x\in \mathcal{C}(V)$.
\item[{\rm (ii)}] $\hat{\psi}(0)\ge \hat{\psi}(p)$   for all  $p\in \hat{\mathcal{C}}(V)$.
\end{itemize}
\end{Thm}

Taking the above theorem into consideration, 
we  define 
\begin{align}
\delta \psi(x)=\sqrt{\psi(0)^2-\psi(x)^2}.
\end{align}
\begin{rem}
{\rm 
\begin{itemize}
\item[1.] Let us consider the hydrogen-like atom discussed in Example \ref{Yukawa}: $V(x)=2^{1/2}/|x|$.
Then we can confirm that all  assumptions in Theorem \ref{PPGS1} are satisfied.
\item[2.]
Using \cite[Theorem 11.7]{LiebLoss}, we see that $0\in \mathcal{C}(V)$ under additional assumptions on $V$. \footnote{
For example, suppose that $V\in L^1(B_r)$, where $B_r=\{x\in \BbbR^d\, |\, |x| <r\}$.
Suppose that $d\ge 2$. If $V\in L^p(B_r)$
for $d\ge p >d/2$, then for all $\alpha<2-d/p$, 
\begin{align}
|\psi(x)-\psi(y)| \le C|x-y|^{\alpha}
\end{align}
for some $C>0$ and all $x, y\in B_{r'}$ with $r'<r$. Hence, $\mathcal{C}(V) \supseteq B_{r'}$ in this case.
Note that all potential given in Example \ref{SchwartzEx} fulfill the assmptions.
}
We can also apply \cite[Theorem C.1.1]{Simon} to check that every potential given in Example \ref{SchwartzEx} satisfies this condition.
$\diamondsuit$
\end{itemize}
}
\end{rem}

\begin{Thm}\label{PPGS2}
Assume that $V^{(1)}$ and $V^{(2)}$ are regular. 
Assume that   $0\in \mathcal{C}(V^{(1)}) \cap \mathcal{C}(V^{(2)})$.
If $V^{(1)}\succeq V^{(2)}$, then we have the following:
\begin{itemize}
\item[{\rm (i)}] $\psi^{(1)}(0) \ge \psi^{(2)}(0)$.
\item[{\rm (ii)}] $\delta \psi^{(1)}(x) \ge \delta \psi^{(2)}(x)$  for all  $x\in 
 \mathcal{C}(V^{(1)}) \cap \mathcal{C}(V^{(2)})
$.
\end{itemize}
\end{Thm}

Next, we define
\begin{align}
\delta \hat{\psi}(p)=\sqrt{\hat{\psi}(0)^2-\hat{\psi}(p)^2}.
\end{align}

\begin{Thm}\label{PPGS3}
Assume that $V^{(1)}$ and $V^{(2)}$ are regular.  Assume that  $0\in  \hat{\mathcal{C}}(V^{(1)}) \cap \hat{\mathcal{C}}(V^{(2)})$. In addition, assume that $\hat{\psi}^{(1)}$ and $\hat{\psi}^{(2)}$ are bounded.
If $V^{(1)}\succeq V^{(2)}$, then we have the following:
\begin{itemize}
\item[{\rm (i)}] $\hat{\psi}^{(1)}(0) \le \hat{\psi}^{(2)}(0)$.
\item[{\rm (ii)}] $\delta \hat{\psi}^{(1)}(p) \le \delta \hat{\psi}^{(2)}(p)$ for all  $p\in \hat{\mathcal{C}}(V^{(1)}) \cap \hat{\mathcal{C}}(V^{(2)}) $.
\end{itemize}
\end{Thm}

\begin{example}
{\rm 
Let $W$ be a regular potential given in Example \ref{Potential1}.
Let $\psi_{\lambda}$ be the unique ground state of
 $H_{\lambda}=-\Delta_x-\lambda W$. 
 For simplicity, we assume that $\psi_{\lambda}$ and $\hat{\psi}_{\lambda}$ are  continuous on $\BbbR^d$ for all $\lambda\in I$.
 We have the following:
 \begin{itemize}
 \item[(i)] $\delta \psi_{\lambda}(x)$ is monotonically increasing in $\lambda$ for all  $x$.
 \item[(ii)] $\delta \hat{\psi}_{\lambda}(p)$ is monotonically decreasing  in $\lambda$ for all $p$.
 \end{itemize}
 Roughly speaking, (i) and (ii) above  mean that, as $\lambda$ increases the shape of $\psi_{\lambda}(x)$ becomes sharper, while that of $\hat{\psi}_{\lambda}(p)$ becomes blunter.  
 In other words,  the  wave function of the particle   is more localized around the origin in the position space  as $\lambda$ increases, while  in the momentum space, it is delocalized.
 These facts can be regarded as an  expression of the Heisenberg's uncertainty principle. 
 $\diamondsuit$ 
}
\end{example}

\subsection{Application II: Properties of $|\psi(x)|^2$}
Let $\varrho(x)=|\psi(x)|^2$. In the context of quantum mechanics, $\varrho(x)$ is interpreted as 
the probability density that the particle is at $x$.
We can apply  the  correlation inequalities to  investigate
properties of $\varrho(x)$.
Here, we present  some examples of applications.

Since $\varrho\in L^1(\BbbR^d; dx)$, $\hat{\varrho}(p)$
exists for all $p\in \BbbR^d$ and is continuous in $p$.

In Section \ref{8}, we  prove the following three theorems:
\begin{Thm}\label{App1}
Assume that $V$ is regular.
\begin{itemize}
\item[{\rm (i)}] $0 < \hat{\varrho}(p)$ for all  $p$.
\item[{\rm (ii)}] $\hat{\varrho}(p)\le \hat{\varrho}(0)=(2\pi)^{-d/2}$ for all $p$. There is equality  if and only if
	     $p=0$.
\item[{\rm (iii)}] $\displaystyle 
 (2\pi)^{d/2}\hat{\varrho}(p) \hat{\varrho}(p') \le
	     \frac{1}{2}\hat{\varrho}(p-p')+\frac{1}{2}\hat{\varrho}(p+p')$
for all  $p, p'$.
\end{itemize} 
\end{Thm} 
Theorem \ref{App1} provides information about the shape of the function $\hat{\varrho}(p)$.

Let $\varrho_n(x)=|\psi_n(x)|^2$.
By the assumption (i) of {\bf (B)}, we readily confirm that $\hat{\varrho}_n(p)$ converges to $\hat{\varrho}(p)$ as $n\to \infty$. The correlation inequalities stated  in this section enable us to obtain  further
information on the convergence:
\begin{Thm}\label{App2}
Assume that $V$ is regular. Then,
$\hat{\varrho}_n(p)$ is monotonically increasing in $n$ for all  $p\in \BbbR^d$.
\end{Thm}

\begin{Thm}\label{App3}
Assume that $V^{(1)}$ and $V^{(2)}$ are regular,  and that
 $V^{(1)}\succeq V^{(2)}$.
Let $\varrho^{(1)}(x)=|\psi^{(1)}(x)|^2$ and $\varrho^{(2)}(x)=|\psi^{(2)}(x)|^2$.
Then, 
$\hat{\varrho}^{(1)}(p) \ge
 \hat{\varrho}^{(2)}(p)$ for all $p\in \BbbR^d$.
\end{Thm} 

Theorem \ref{App3} suggests that, as the strength of $V$ becomes larger, the probability 
density has a tendency to localized around the origin in the position space.
This result is consistent with the results in Section \ref{SecApp1}.
 
\begin{example}
{\rm 
Let $W$ be a regular potential given in Example \ref{Potential1}.
Let $\psi_{\lambda}$ be the unique ground state of
 $H_{\lambda}=-\Delta_x-\lambda W$, and let
 $\varrho_{\lambda}(x)=|\psi_{\lambda}(x)|^2$. Then, $\hat{\varrho}_{\lambda}(p)$
is monotonically increasing in $\lambda\in I$ for all $p\in \BbbR^d$.
$\diamondsuit$ 
}
\end{example}

\section{Preliminaries}\label{3}
\setcounter{equation}{0}
In order to prove the theorems in Section \ref{Sec2}, we must introduce several operator inequalities associated with self-dual cones.
\subsection{Self-dual cones}
Let $\h$ be a complex Hilbert space.
By a {\it convex  cone}, we understand a closed convex set  $\Cone\subset \h$
such that $t\Cone \subseteq \Cone$ for all $t\ge 0$ and $\Cone\cap (-\Cone)=\{0\}$. In what follows, we always assume that $\Cone\neq
\{0\}$.

\begin{define}{\rm
The {\it dual cone   of} $\Cone$ is defined by 
\begin{align}
\Cone^{\dagger}=\{\eta\in \h\, |\, \la \eta|\xi\ra\ge 0\ \forall \xi\in
\Cone \}.
\end{align} 
 We say that $\Cone$ is {\it self-dual} if 
$
\Cone=\Cone^{\dagger}. 
$ $\diamondsuit$
}\end{define}

\begin{define}[\cite{Faris}]\label{HilCone}
{\rm
Let $\h$ be a complex Hilbert space.  A convex cone
 $\Cone$ in $\h$ is called  a {\it Hilbert cone}, if it
 satisfies
 the following: 
\begin{itemize}
\item[(i)] $ \la \xi| \eta\ra\ge 0$ for all $\xi, \eta\in \Cone$.
\item[(ii)] Let $\h_{\BbbR}$
 be a real closed subspace of $\h$ generated by $\Cone$ . Then
	     for all $\xi\in \h_{\BbbR}$, there exist $\xi_+,
	     \xi_-\in \Cone$ such that $\xi=\xi_+-\xi_-$ and $\la \xi_+|
	     \xi_-\ra=0$.
\item[(iii)] $\h=\h_{\BbbR}+i
	    \h_{\mathbb{R}}= \{\xi+ i \eta\, |\, \xi, \eta\in
	     \h_{\BbbR}\}$.\ \ \ $\diamondsuit$
\end{itemize} 
}
\end{define} 

\begin{rem}\label{DecRI}
{\rm
Let $\Cone$ be a Hilbert cone in $\h$. For each $\xi\in \h$,  we have the following expression:
\begin{align}
\xi=(\xi_1-\xi_2)+i(\xi_3-\xi_4), \label{ReIm}
\end{align}
where $\xi_1, \xi_2, \xi_3$ and $\xi_4$ satisfy $\xi_1, \xi_2, \xi_3,
 \xi_4\in \Cone$, $\la \xi_1|\xi_2\ra=0$ and $\la \xi_3|\xi_4\ra=0$. $\diamondsuit$
}
\end{rem}

\begin{Thm}\label{SAH}
Let $\Cone$ be a convex cone in $\h$.
The following are equivalent:
\begin{itemize}
\item[{\rm (i)}]$\Cone$ is a  self-dual cone.
\item[{\rm (ii)}] $\Cone$ is a Hilbert cone. 
\end{itemize} 
\end{Thm} 
{\it Proof.} For (i) $\Rightarrow$ (ii),  see, e.g.,  \cite{Bos} or \cite[Proof of Proposition 2.58]{BR1}.

Suppose that  $\Cone$ is a Hilbert cone.  We easily  see that 
$\Cone\subseteq \Cone^{\dagger}$  by Definition \ref{HilCone} (i).
We will show the inverse. Let $\xi\in \Cone^{\dagger}$.
By (\ref{ReIm}), we can write $\xi$ as 
$\xi=(\xi_{R, +}-\xi_{R, -})+i(\xi_{I, +}-\xi_{I, -})$ with 
$\xi_{R, \pm}, \xi_{I, \pm} \in \Cone$, $\la \xi_{R, +}|\xi_{R, -}\ra=0$
and $\la \xi_{I, +}|\xi_{I, -}\ra=0$. 
Assume that $\xi_{I, +}\neq 0$. Then $\la \xi|\xi_{I, +}\ra$ is a
complex number, which contradicts with the fact that $\la \xi|\eta\ra\ge
0$ for all $\eta\in \Cone$. Thus, $\xi_{I, +}=0$.
Similarly, we have $\xi_{I, -}=0$.
Next, assume that $\xi_{R, -}\neq 0$. Because $\xi_{R, -}\in \Cone$, we
have
\begin{align}
0\le \la \xi|\xi_{R, -}\ra=-\|\xi_{R, -}\|^2<0,
\end{align} 
which is a contradiction. Hence, we conclude that $\xi=\xi_{R, +}\in
\Cone$. $\Box$

\begin{define}
{\rm 
\begin{itemize}
\item A vector $\xi$ is said to be  {\it positive w.r.t. $\Cone$} if $\xi\in
 \Cone$.  We write this as $\xi\ge 0$  w.r.t. $\Cone$.

 \item A vector $\eta\in \Cone$ is called {\it strictly positive
w.r.t. $\Cone$} whenever $\la \xi| \eta\ra>0$ for all $\xi\in
\Cone\backslash \{0\}$. We write this as $\eta>0 $
w.r.t. $\Cone$. $\diamondsuit$

\end{itemize} 
}
\end{define} 
\begin{example}\label{L2+Ex}
{\rm 
For each $d\in \BbbN$, we  set 
\begin{align}
L^2(\BbbR^d; du)_+=\{f\in L^2(\BbbR^d; du)\, |\, f(u)\ge 0\ \ \mbox{a.e. $u$}
\}.
\end{align}
$L^2(\BbbR^d;  du)_+$ is a self-dual cone in $L^2(\BbbR^d; du)$.
$f\ge 0$ w.r.t. $L^2(\BbbR^d; du)_+$
 if and only if $f(u) \ge 0$ a.e. $u$. On the other hand, $f >0$
 w.r.t. $L^2(\BbbR^d; du)_+$ if and only if $f(u)>0$ a.e. $u$.
 $\diamondsuit$ 
}
\end{example} 

\subsection{Operator inequalities associated with self-dual cones}

In  subsequent  sections, we  use the following operator inequalities.
\begin{define}{\rm 
Let $A, B\in \mathscr{B}(\h)$. Let $\Cone$ be a self-dual cone in $\h$.

 If $A \Cone\subseteq \Cone,$\footnote{
For each subset $\mathfrak{C}\subseteq \h$, $A\mathfrak{C}$ is
	     defined by $A\mathfrak{C}=\{Ax\, |\, x\in \mathfrak{C}\}$.
} we then 
write  this as  $A \unrhd 0$ w.r.t. $\Cone$.\footnote{This
 symbol was introduced by Miura \cite{Miura}, see also \cite{KiRo2}.} In
	     this case, we say that {\it $A$ preserves the
positivity w.r.t. $\Cone$.}  Suppose that $A\h_{\BbbR}\subseteq
 \h_{\BbbR}$ and $B\h_{\BbbR} \subseteq
	     \h_{\BbbR}$. If $(A-B) \Cone\subseteq
	     \Cone$, then we write this as $A \unrhd B$ w.r.t. $\Cone$. $\diamondsuit$
} 
\end{define} 

\begin{rem}\label{Pequiv}
{\rm
\begin{itemize}
\item[(i)] $A\unrhd 0$ w.r.t. $\Cone$ $\Longleftrightarrow$ $ \la \xi |A\eta\ra\ge
 0$
for all $\xi, \eta\in \Cone$. 
\item[(ii)]
Let $A\in \mathscr{B}(\h)$. If $A\h_{\BbbR} \subseteq \h_{\BbbR}$, we say that $A$
{\it preserves the reality w.r.t. $\Cone$}.
The following fact will be often used: if $A$ preserves the positivity w.r.t. $\Cone$, then 
$A$ preserves the reality w.r.t. $\Cone$.
$\diamondsuit$
\end{itemize}
}
\end{rem}

The following proposition is fundamental to this paper.

\begin{Prop}\label{Miura}
Let $A, B, C, D\in \mathscr{B}(\h)$ and let $a, b\in
 \BbbR$. 
\begin{itemize}
\item[{\rm (i)}] If $A\unrhd 0, B\unrhd 0$ w.r.t. $\Cone$ and
	     $a, b\ge 0$, then $aA +bB \unrhd 0$
	     w.r.t. $\Cone$.
\item[{\rm (ii)}] If $A \unrhd B \unrhd 0$ and $C\unrhd D \unrhd 0$
	     w.r.t. $\Cone$,
	     then
$AC\unrhd BD \unrhd 0$ w.r.t. $\Cone$.
\item[{\rm (iii)}] If $A \unrhd 0 $ w.r.t. $\Cone$, then $A^*\unrhd 0$ w.r.t. $\Cone$.
\end{itemize} 
\end{Prop} 
{\it Proof.} (i) is trivial.

(ii) If $X\unrhd 0$ and $Y\unrhd 0$
w.r.t. $\Cone$, 
we have $XY\Cone\subseteq X\Cone \subseteq \Cone$.
Hence,  it holds that $XY\unrhd 0$
w.r.t. $\Cone$.
Hence, we have 
\begin{align*}
AC-BD=\underbrace{A}_{\unrhd 0}\underbrace{(C-D)}_{\unrhd 0}+
 \underbrace{(A-B)}_{\unrhd 0} \underbrace{D}_{\unrhd 0} \unrhd 0\ \ \
 \mbox{w.r.t. $\Cone$}.
\end{align*} 

(iii) For each $\xi, \eta\in \Cone$, we know that 
\begin{align}
\la \xi|A^*\eta\ra=\la \underbrace{A}_{\unrhd 0} \underbrace{\xi}_{\ge 0}|
\underbrace{\eta}_{\ge 0}\ra \ge 0.
\end{align}
Thus, by Remark \ref{Pequiv} (i), we conclude (iii).  $\Box$
\medskip\\

\begin{define}
{\rm 
Let $A\in \mathscr{B}(\h)$.
We write  $A\rhd 0$ w.r.t. $\Cone$, if  $A\xi >0$ w.r.t. $\Cone$ for all $\xi\in
\Cone \backslash \{0\}$. 
 In this case, we say that {\it $A$ improves the
positivity w.r.t. $\Cone$.} $\diamondsuit$
}
\end{define}

\begin{define}\label{DefErg}
{\rm 
Let $A\in \mathscr{B}(\h)$.
Assume that $A\unrhd 0$ w.r.t. $\Cone$. 
  We say that $A$ is {\it ergodic  w.r.t.} $\Cone$
 if for each $\xi, \eta\in \Cone \backslash \{0\}$, there exists an
 $n\in \{0\} \cup \BbbN$ such that $\la \xi|A^n\eta \ra>0$. 
Note  that the number $n$ could depend on $\xi$ and $\eta$. 
$\diamondsuit$
}
\end{define}

\subsection{A canonical cone in $\mathscr{L}^2(\h )$}

Let $\h $ be a complex Hilbert space. The set of all
Hilbert--Schmidt class operators on $\h $ is denoted  by
$\mathscr{L}^2(\h )$, i.e.,  
$
\mathscr{L}^2(\h )=\{
\xi\in \mathscr{B}(\h )\, |\, \Tr[\xi^* \xi]<\infty
\}$.
$\mathscr{L}^2(\h)$ is a two-sided ideal in $\mathscr{B}(\h)$.
Henceforth, we regard $\mathscr{L}^2(\h )$ as a Hilbert space equipped
with  the inner product $\la \xi| \eta \ra_{\mathscr{L}^2}=\Tr[\xi^*
\eta],\,   \xi, \eta\in \mathscr{L}^2(\h )$. 
\begin{define}{\rm 
For each $A\in \mathscr{B}(\h )$, the {\it left multiplication
operator} is defined by
\begin{align}
\mathcal{L}(A)\xi=A\xi,\ \ \xi\in \mathscr{L}^2(\h ).
\end{align} 
Similarly, the {\it right multiplication operator} is defined by 
\begin{align}
\mathcal{R}(A)\xi=\xi A, \ \ \xi\in \mathscr{L}^2(\h ).
\end{align} 
Note that $\mathcal{L}(A)$ and
 $\mathcal{R}(A)$
belong to $\mathscr{B}(\mathscr{L}^2(\h))$, where $\mathscr{B}(\mathscr{L}^2(\h))$
 is the set of all bounded linear operators on $\mathscr{L}^2(\h)$. $\diamondsuit$
}
\end{define} 

It is not  difficult to check that 
\begin{align}
\mathcal{L}(A)\mathcal{L}(B)=\mathcal{L}(AB),\ \
 \mathcal{R}(A)\mathcal{R}(B)=\mathcal{R}(BA),\ \ A, B\in \mathscr{B}(\h ).
\end{align}

 Let $\vartheta$ be an
antiunitary  operator on $\h$.\footnote{
We say that a bijective map  $\vartheta$  on $\h$ is  {\it antiunitary}
 if $\la \vartheta x|\vartheta y\ra=\overline{\la x|y\ra}$ for all $x,
 y\in\h$.
} 
Let $\Phi_{\vartheta}$ be an isometric
isomorphism  from $\mathscr{L}^2(\h)$ onto $\h\otimes \h$ defined by
\begin{align}
\Phi_{\vartheta}(|x\ra\la y|)=x\otimes \vartheta y\ \ \ \forall x,y\in \h,
\end{align} 
where the linear operator $|x\ra\la y|$ is defined by $|x\ra\la y|z=\la y|z\ra x$ for all $z\in \h$.
Then,
\begin{align}
\mathcal{L}(A) =\Phi_{\vartheta}^{-1} A\otimes \one\Phi_{\vartheta} ,\ \
 \ \mathcal{R}(\vartheta
 A^*\vartheta)=\Phi_{\vartheta}^{-1} \one \otimes A\Phi_{\vartheta} \label{Bare}
\end{align} 
for each $A\in \mathscr{B}(\h)$. We  write these facts simply  as 
\begin{align}
\h\otimes \h=\mathscr{L}^2(\h),\ \ A\otimes \one =\mathcal{L}(A),\ \ \one \otimes
 A=\mathcal{R}(\vartheta A^*\vartheta), \label{Ident}
\end{align} 
if no confusion arises.

The left and right multiplication operators can be extended to unbounded operators by
(\ref{Bare}) as follows.
Let $A$ be  a densely defined closed operator on $\h$.
The left multiplication operator $\mathcal{L}(A)$ and 
the right multiplication operator $\mathcal{R}(A)$  are defined as 
$
\mathcal{L}(A) =\Phi_{\vartheta}^{-1} A\otimes \one\Phi_{\vartheta}
$
and  
$
\mathcal{R}(A) =\Phi_{\vartheta}^{-1} \one\otimes \vartheta A^* \vartheta \Phi_{\vartheta}
$, respectively. 

\begin{rem}{\rm
\begin{itemize}
\item[{\rm (i)}]Both $\mathcal{L}(A)$ and $\mathcal{R}(A)$ are closed
 operators on $\mathscr{L}^2(\h)$. 
\item[{\rm (ii)}]If $A$ is self-adjoint,    then
 $\mathcal{L}(A)$ and $\mathcal{R}(A)$ are self-adjoint. 
\item[{\rm (iii)}] We will also use the conventional identification
	     (\ref{Ident}). $\diamondsuit$ 
\end{itemize} 
}
\end{rem} 

Recall that a  linear operator $A$ on $\h$ is said to be {\it positive} if $\la \xi|
A \xi\ra_{\h} \ge 0$ for all $\xi\in \h$. We write this as $A\ge 0$.

\begin{define}{\rm 
A canonical   cone in $\mathscr{L}^2(\h )$ is given by
\begin{align}
\mathscr{L}^2(\h )_+= \Big\{\xi\in \mathscr{L}^2(\h )\, \Big|\,\mbox{$\xi$ is
 self-adjoint and $\xi\ge 0$
  as
 an operator on $\h $} \Big\}.\ \ \ \diamondsuit
\end{align} 
}
\end{define} 

\begin{Thm}\label{SDL2}
$\mathscr{L}^2(\h)_+$ is a self-dual cone in $\mathscr{L}^2(\h)$.
\end{Thm}  
{\it Proof.}
 We now  check the   conditions (i)--(iii) in Definition \ref{HilCone}.

(i) Let $\xi, \eta\in \mathscr{L}^2(\h)_+$. Since $\xi^{1/2} \eta \xi^{1/2}\ge
0$, we have $\la \xi| \eta\ra_{\mathscr{L}^2}=\Tr[\xi\eta]=\Tr[\xi^{1/2}\eta
\xi^{1/2}] \ge 0$.

(ii) Note that $\mathscr{L}^2(\h)_{\BbbR}=\{\xi\in \mathscr{L}^2(\h)\, |\, \mbox{$\xi$ is
self-adjoint }\}$. Let $\xi\in \mathscr{L}^2(\h)_{\BbbR}$. By the spectral
theorem,
there  is a projection valued measure $\{E(\cdot)\}$ such that
$\xi=\int_{\BbbR} \lambda dE(\lambda)$.
Denote $
\xi_+=\int_0^{\infty} \lambda dE(\lambda)
$
and 
$
\xi_-=\int_{-\infty}^0 (-\lambda)dE(\lambda)
$. Clearly, it holds that $\xi_+\xi_-=0, \xi_{\pm}\in \mathscr{L}^2(\h)_+$ and
$\xi=\xi_+-\xi_-$. Thus, (ii) is satisfied.

(iii) For each $\xi\in \mathscr{L}^2(\h)$, we have $\xi=\xi_{R}+i \xi_I$, where 
$\xi_R=(\xi+\xi^*)/2$ and  $\xi_I=(\xi-\xi^*)/2i$. Trivially, $\xi_R,
\xi_I\in \mathscr{L}^2(\h)_{\BbbR}$. 
Hence, $\mathscr{L}^2(\h)_+$ is a Hilbert cone. By Theorem \ref{SAH}, we
conclude that $\mathscr{L}^2(\h)_+$ is a self-dual cone.  $\Box$

\begin{Prop}\label{GeneralPP}
Let $A\in \mathscr{B}(\h )$. We have 
 $\mathcal{L}(A^*)\mathcal{R}(A)\unrhd 0$ w.r.t. $\mathscr{L}^2(\h )_+$.
\end{Prop} 
{\it Proof.} For each $\xi\in \mathscr{L}^2(\h )_+$,  we have 
$
\mathcal{L}(A^*)\mathcal{R}(A)\xi=A^*\xi A \ge 0.
$ $\Box$

\begin{rem}
{\rm 
As we noted in references \cite{Miyao6, Miyao8}, Proposition \ref{GeneralPP} is closely
 related
to  spin reflection positivity \cite{Lieb}; see also 
references \cite{Faris,Gross}. $\diamondsuit$
}
\end{rem}

\section{Proof of Proposition \ref{GSUni} and Theorem \ref{FirstInq}} \label{4}
\setcounter{equation}{0}

\subsection{Proof of Proposition \ref{GSUni}}
Let $\mathcal{F}$ be the Fourier transformation on $L^2(\BbbR^d; dx)$:
\begin{align}
(\mathcal{F}f)(p)=(2\pi)^{-d/2} \int_{\BbbR^d} e^{-i p\cdot x} f(x) dx,\ \ \ f\in L^2(\BbbR^d; dx).
\end{align}
It is known that $\mathcal{F}$ is a unitary operator from $L^2(\BbbR^d; dx)$ onto $L^2(\BbbR^d; dp)$.

Let $H_n=-\Delta_x-V_n$ and let
 $\hat{H}_n=\mathcal{F} H_n\mathcal{F}^{-1}$. We have
\begin{align}
\hat{H}_n&=p^2- V_n(-i \nabla_p),
\end{align} 
where $p^2$ stands for the multiplication operator.
Of course, $\hat{H}_n$ acts on $L^2(\BbbR^d; dp)$.

\begin{rem}
{\rm 
By the condition {\bf (B)}, $\hat{V}_n\in L^1(\BbbR^d; dp)$, which implies that $V_n\in L^{\infty}(\BbbR^d; dx)$. Thus, the linear operator $V_n(-i\nabla_p)$ can be defined by (\ref{DefNabla}). $\diamondsuit$
}
\end{rem}

\begin{lemm}\label{SchB}
For all $n\in \BbbN$, we have the following:
\begin{itemize}
\item[{\rm (i)}]$V_n(-i \nabla_p) \unrhd 0$ w.r.t. $L^2(\BbbR^d; dp)_+$, where $L^2(\BbbR^d; dp)_+$ is defined in Example \ref{L2+Ex}.
\item[{\rm (ii)}] 
$\exp(-\beta \hat{H}_n)\unrhd 0$
	     w.r.t. $L^2(\BbbR^d; dp)_+$ for all $\beta \ge 0$.
\end{itemize} 
\end{lemm} 
{\it Proof.}
Let 
$\nabla_p=(D_{p_1},\dots, D_{p_d})$, where   $D_{p_j}$  is   the
(generalized) differential operator on $L^2(\BbbR^d; dp)$.

(i) Since $e^{i k\cdot (-i \nabla_p)}$ is a translation, we see that 
$e^{ik\cdot (-i\nabla_p)} \unrhd 0$ w.r.t. $L^2(\BbbR^d;  dp)_+$ for all
$k\in \BbbR^d$. 
Thus, by (ii) of {\bf (B)} and the fact $\mathcal{F} e^{ik\cdot x}
\mathcal{F}^{-1}
=
e^{ik \cdot (-i \nabla_{p})}$
, we have 
\begin{align}
V_n(-i \nabla_p)
=(2\pi)^{-d/2}\int_{\BbbR^d} \underbrace{e^{ik\cdot (-i\nabla_p)}}_{\unrhd 0}
 \underbrace{\hat{V}_n(k)}_{\ge 0}dk \unrhd 0\ \ \  \mbox{w.r.t. $L^2(\BbbR^d; dp)_+$.}
\end{align}

(ii) We know that the multiplication operator $\ex^{-\beta p^2}$
satisfies $\ex^{-\beta p^2}\unrhd 0$ w.r.t. $L^2(\BbbR^d; dp)_+$.
Thus,  applying Theorem \ref{StandPP}, we conclude (ii). $\Box$
\medskip\\

Before we proceed, we take note of  the following fact.

\begin{lemm}\label{SetErg}
Let $\mathbb{B}^d$ be the Borel algebra on $\BbbR^d$.
Let $B_1, B_2\in \mathbb{B}^d$ with $|B_1|>0$ and $|B_2|>0$, where
 $|\cdot|$
is the Lebesgue measure.
For any $\vepsilon>0$, we set 
\begin{align}
\mathcal{S}_{\vepsilon}^{(\ell)}=\Big\{
(p, p_1, \dots, p_{\ell})\in \BbbR^{d\times (\ell+1)}\, \Big|
\, p\in B_2,\, p+p_1+\cdots+p_{\ell}\in B_1,\, p_1, \dots, p_{\ell}\in B_{\vepsilon}(0)
\Big\}.
\end{align} 
Then, for each $\vepsilon>0$, there exists an $\ell \in \BbbN_0:=\{0\} \cup \BbbN$  such that 
$\big|\mathcal{S}_{\vepsilon}^{(\ell)}\big|>0$.
\end{lemm} 
{\it Proof.} Without loss of generality, we may  assume that $B_1$ and
$B_2$ are connected sets.
For each $p_1, \dots, p_{\ell}\in \BbbR^d$ and $\vepsilon>0$, we set
\begin{align}
\mathcal{S}_{\vepsilon}^{(\ell)}(p_1, \dots, p_{\ell})
=\Big\{
p\in \BbbR^d\, \Big|\, p\in B_2,\, p+p_1+\cdots+p_{\ell} \in B_1
\Big\}.
\end{align} 
Note that $\mathcal{S}_{\vepsilon}^{(\ell)}(p_1, \dots, p_{\ell})$ could
be empty. For each $\vepsilon>0$, there exist an $\ell\in \BbbN_0$ and
$p_1, \dots, p_{\ell} \in B_{\vepsilon}(0)$ such that 
$
\Big|
B_2\cap (B_1-p_1-\cdots-p_{\ell})
\Big|>0
$, where 
$
B_1-p_1-\cdots-p_{\ell}=\{
p-p_1-\cdots-p_{\ell}\, |\, p\in B_1
\}
$. Thus, for these $\ell\in \BbbN_0$ and $p_1, \dots, p_{\ell} \in B_{\vepsilon}(0)$, $
\big|\mathcal{S}_{\vepsilon}^{(\ell)}(p_1, \dots, p_{\ell})\big|>0
$. Because  $
\big|\mathcal{S}_{\vepsilon}^{(\ell)}(p_1, \dots, p_{\ell})\big|
$  is continuous in $p_1, \dots, p_{\ell}$, we have 
\begin{align}
\big|\mathcal{S}_{\vepsilon}^{(\ell)}\big|
=\int_{
(B_{\vepsilon}(0))^{\times \ell}} dp_1\cdots dp_{\ell}
\big|\mathcal{S}_{\vepsilon}^{(\ell)}(p_1, \dots, p_{\ell})\big|
>0.
\end{align} 
This completes the proof. $\Box$

\begin{Prop}\label{SchErg}
For each $n\in \BbbN$,
$V_n(-i \nabla_p)$ is ergodic w.r.t. $L^2(\BbbR^d; dp)_+$ (see Definition \ref{DefErg}).
\end{Prop} 
{\it Proof.}
Recall that, by (ii) of the assumption {\bf (B)}, there exists an
$\vepsilon>0$ such that $\mathrm{supp} \hat{V}_n \supset B_{\vepsilon}(0)$.

 Let $f_1, f_2\in L^2(\BbbR^d;  dp)_+\backslash \{0\}$.
Because $f_1$ and $f_2$ are non-zero, there
exist   $B_1, B_2\in \mathbb{B}^d$ such that  $|B_1|>0,\  |B_2|>0$, and  
$f_1(p)>0$ on $ B_1$, $f_2(p)>0$ on $ B_2$.
By Lemma \ref{SetErg}, there exists an $\ell\in \BbbN_0$ such that 
$\big|
\mathcal{S}_{\vepsilon}^{(\ell)}
\big|>0$. In addition, we have 
\begin{align}
f_2(p) \Big(
e^{i (p_1+\cdots+p_{\ell})\cdot (-i \nabla_p)}
f_1
\Big)(p)
=f_2(p)f_1(p+p_1+\cdots+p_{\ell})>0
\end{align} 
for all $p, p_1, \dots, p_{\ell}\in \BbbR^d$ such that 
$(p, p_1, \dots, p_{\ell}) \in \mathcal{S}_{\vepsilon}^{(\ell)}$.
Therefore, we obtain
\begin{align}
&\la f_2| V_n^{\ell}(-i\nabla_p) f_1\ra
 \no
=&(2\pi)^{-nd/2}\int_{\BbbR^d}dp \int_{(\BbbR^d)^{\times \ell}} dp_1\cdots
 dp_{\ell}
\underbrace{\hat{V}_n(p_1)\cdots \hat{V}_{n}(p_{\ell})}_{\ge 0}
\underbrace{f_2(p)f_1(p+p_1+\cdots +p_{\ell})}_{\ge 0}\no
\ge &
(2\pi)^{-nd/2}\int_{
\mathcal{S}_{\vepsilon}^{(\ell)}
} dp dp_1\cdots
 dp_{\ell}
\underbrace{\hat{V}_n(p_1)\cdots \hat{V}_{n}(p_{\ell})}_{>0}
\underbrace{f_2(p) f_1(p+p_1+\cdots +p_{\ell})}_{>0}\no
>&0.
\end{align} 
This completes the proof. $\Box$

\begin{Prop}\label{PI}
 We have $\exp(-\beta \hat{H})\rhd 0$
 w.r.t. $L^2(\BbbR^d;  dp)_+$ for all $\beta >0$.
\end{Prop} 
{\it Proof.}
By Lemma \ref{SchB} (ii), Theorem \ref{ErgPI} and Proposition \ref{SchErg}, we have $\exp(-\beta \hat{H}_{n}) \rhd 0$
w.r.t. $L^2(\BbbR^d; dp)_+$ for all $\beta >0$ and $n\in \BbbN$.

For each $m,n\in \BbbN$ with $n\ge m$,   we have, by the assumption (iii) of {\bf (B)},
\begin{align}
V_n(-i \nabla_p)-V_m(-i \nabla_p)=(2\pi)^{-d/2}\int_{\BbbR^d}
\underbrace{(\hat{V}_n(k)-\hat{V}_m(k))}_{\ge 0}
 \underbrace{e^{i k \cdot(-i
 \nabla_p)}}_{\unrhd 0} dk
 \unrhd 0
\end{align} 
w.r.t. $L^2(\BbbR^d; dp)_+$.
By Theorem \ref{Mono}, we obtain that 
$
\exp(-\beta \hat{H}_{n}) \unrhd \exp(-\beta
\hat{H}_{m})
$
w.r.t. $L^2(\BbbR^d; dp)_+$ for all $\beta \ge 0$. Taking $n\to \infty$,
we conclude that  $
\exp(-\beta \hat{H}) \unrhd \exp(-\beta
\hat{H}_{m})
$
w.r.t. $L^2(\BbbR^d; dp)_+$ for all $\beta \ge 0$, where 
$\hat{H}=\mathcal{F} H\mathcal{F}^{-1}$. 
Since  $\exp(-\beta \hat{H}_m) \rhd
0$ w.r.t. $L^2(\BbbR^d; dp)_+$ for all $\beta>0$, we finally arrive at 
\begin{align}
\exp(-\beta \hat{H}) \unrhd \exp(-\beta
\hat{H}_{m})\rhd 0\ \ \mbox{w.r.t. $L^2(\BbbR^d; dp)_+$ for
 all $\beta >0$}.
\end{align} 
Thus we are done. $\Box$

\begin{flushleft}
{\it Proof of Proposition \ref{GSUni}}
\end{flushleft} 
It is well-known that  $\exp(-\beta H) \rhd 0$ and $\exp(-\beta H_n) \rhd
0$ w.r.t. $L^2(\BbbR^d; dx)_+$ for all $\beta >0$, see, e.g.,
\cite[Theorem XIII. 45]{ReSi4}.
Thus, we conclude the uniqueness of ground states by \cite[Theorem XIII. 4.4]{ReSi4}.
Simultaneously, we obtain (i).

By \cite[Theorem XIII. 45]{ReSi4}  and Proposition \ref{PI} , we conclude (ii). $\Box$

\subsection{Proof of Theorem \ref{FirstInq}}

\begin{lemm}\label{AP1}
Let $f\in \mathfrak{A}$.
\begin{itemize}
\item[{\rm (i)}] $\mathcal{F} f\mathcal{F}^{-1} \unrhd 0$
	     w.r.t. $L^2(\BbbR^d; dp)_+$.
\item[{\rm (ii)}] $
f(-i\nabla_x) \unrhd 0
$ w.r.t. $L^2(\BbbR^d; dx)_+$.
\end{itemize} 
\end{lemm} 
{\it Proof.}
(i) 
Because $\mathcal{F} f\mathcal{F}^{-1}=f(-i \nabla_p)$ and 
$\mathcal{F}e^{ik\cdot x} \mathcal{F}^{-1}=
e^{i k\cdot(-i \nabla_p)} \unrhd 0$ w.r.t. $L^2(\BbbR^d; dp)_+$, we
have 
\begin{align}
\mathcal{F}f \mathcal{F}^{-1}= (2\pi)^{-d/2}\int_{\BbbR^d}
 \underbrace{\hat{f}(k)}_{\ge 0}
\underbrace{e^{ik \cdot (-i\nabla _p)}}_{\unrhd 0}dk \unrhd 0\ \ \
 \mbox{w.r.t. 
 $L^2(\BbbR^d; dp)_+$}.
\end{align} 

(ii) Because $e^{i k\cdot (-i \nabla_x)} \unrhd 0$ w.r.t. $L^2(\BbbR^d;
dx)_+$, we have
\begin{align}
f(-i \nabla_x)=(2\pi)^{-d/2} \int_{\BbbR^d} 
\underbrace{\hat{f}(k)}_{\ge 0}
\underbrace{
e^{i k\cdot (-i \nabla_x)}
}_{\unrhd 0} dk \unrhd 0\ \ \mbox{w.r.t. $L^2(\BbbR^d; dx)_+$}.
\end{align} 
This completes the proof. $\Box$
\begin{flushleft}
{\it Proof of Theorem \ref{FirstInq}}
\end{flushleft} 
(i) By Proposition \ref{GSUni} (ii) and Lemma \ref{AP1} (i), 
\begin{align}
\la f\ra=\la \underbrace{\hat{\psi}}_{>0}|
\underbrace{\mathcal{F}f \mathcal{F}^{-1}}_{\unrhd 0}
\underbrace{ \hat{\psi}}_{>0}\ra\ge 0.
\end{align} 
By Theorem \ref{VecP3}, the equality holds if and only if $f=0$.

(ii)  By Proposition \ref{GSUni} (i) and Lemma \ref{AP1} (ii), 
\begin{align}
\la f(-i \nabla_x)\ra=\la \underbrace{\psi}_{>0}|
\underbrace{f (-i\nabla_x)}_{\unrhd 0}
\underbrace{ \psi}_{>0}\ra\ge 0.
\end{align} 
By Theorem \ref{VecP3}, the equality holds if and only if $f=0$.
$\Box$

\subsection{Proof of Remark \ref{RealPRem}}\label{PfRem}
(i) Let $J$ be a natural involution defined by $J\psi=\overline{\psi}$ for each $\psi\in L^2(\BbbR^d; dx)$.
Thus, we have 
\begin{align}
\la J\chi|J\vphi\ra=\overline{\la \chi|\vphi\ra},\ \ \chi,  \vphi\in L^2(\BbbR^d; dx). \label{Cong}
\end{align}
Because $g$ is in $\mathfrak{A}_{\mathrm{e}}$, we have
$\overline{g(x)}=g(-x)=g(x)$, that is, $g$ is real-valued, which implies that $JgJ=g$.
Because $Je^{ik\cdot (-i \nabla_x)} J=e^{ik\cdot (-i\nabla_x)}$, we have
$Jf(-i\nabla_x)J=f(-i\nabla_x)$. Therefore, since $J\psi=\psi$ by Proposition \ref{GSUni} (i), we have
\begin{align}
\overline{\la \psi|f(-i\nabla_x) g\psi\ra}=\la J\psi|Jf(-i\nabla_x) g\psi\ra
=\la J\psi|Jf(-i\nabla_x)JJ gJJ\psi\ra
=\la \psi|f(-i\nabla_x) g\psi\ra.
\end{align}
Thus, we conclude (i).

(ii) First, assume that $g(x) \ge 0$. Then $g\unrhd 0$ w.r.t. $L^2(\BbbR^d; dx)_+$.
Thus, $f(-i\nabla_x) g \unrhd 0$ w.r.t. $L^2(\BbbR^d; dx)_+$ by Lemma \ref{AP1}
 (ii), which implies that 
 \begin{align}
 \la \underbrace{\psi}_{>0} \underbrace{|f(-i\nabla_x) g}_{\unrhd 0} \underbrace{\psi}_{>0}\ra \ge 0.
 \end{align}
Next, assume that $f(x)\ge 0$. Remark that  $f(p) \unrhd 0$ and $g(-i\nabla_p)=\mathcal{F}  g \mathcal{F}^{-1}\unrhd 0$ w.r.t. $L^2(\BbbR^d; dp)_+$ by Lemma \ref{AP1} (i). 
Hence,  by Proposition \ref{GSUni} (ii),
\begin{align}
\la \psi|f(-i\nabla_x) g\psi\ra=\la \underbrace{\hat{\psi}}_{>0}|
\underbrace{f(p) g(-i\nabla_p)}_{\unrhd 0}  \underbrace{\hat{\psi}}_{>0}\ra \ge 0.
\end{align}
Thus we are done. $\Box$

\section{Proof of Theorem \ref{MonoEx}}\label{5}
\setcounter{equation}{0}
\subsection{Extended Hamiltonian }
Consider the  extended Hamiltonian
\begin{align}
\mathbb{H}_n=H_n\otimes \one +\one \otimes H_n
\end{align} 
acting on the  doubled  Hilbert space
$\h _{\mathrm{ext}}:=\h \otimes \h \cong L^2(\BbbR^d\times \BbbR^d; dx_1dx_2)$.

Let us introduce a new coordinate system $(X_1, X_2)$ by 
\begin{align}
X_1=\frac{x_2-x_1}{\sqrt{2}},\ \ \ X_2= \frac{x_2+x_1}{\sqrt{2}}.\label{Coordinate}
\end{align} 
Trivially, 
\begin{align}
\nabla_{x_1}=-\frac{1}{\sqrt{2}} \nabla_{X_1}+\frac{1}{\sqrt{2}}\nabla_{X_2},\ \ \ 
\nabla_{x_2}=\frac{1}{\sqrt{2}}\nabla_{X_1}+\frac{1}{\sqrt{2}}\nabla_{X_2}, \label{Coordinate2}
\end{align}
implying 
\begin{align}
-\Delta_{x_1}-\Delta_{x_2}=-\Delta_{X_1}-\Delta_{X_2}.
\label{Delta}
\end{align}

We define an antiunitary operator $\vartheta$ on $L^2(\BbbR^d; dX)$ by 
\begin{align}
(\vartheta \phi)(X)=\overline{\phi(X)}\ \ \mbox{a.e. $X$} \label{Concrete}
\end{align} 
for each $\phi\in L^2(\BbbR^d; dX)$. Using $\vartheta$, we obtain the
following identifications:
\begin{align}
\h _{\mathrm{ext}}
&= L^2(\BbbR^d; dx) \otimes L^2(\BbbR^d; dx)\no
&\cong L^2(\BbbR^d\times \BbbR^d; dx_1dx_2 )\no
&= L^2(\BbbR^d\times \BbbR^d; dX_1 dX_2)\no
&\cong L^2(\BbbR^d; dX)\otimes L^2(\BbbR^d; dX)\no
&=\mathscr{L}^2(L^2(\BbbR^d;dX)). \label{Double}
\end{align} 
In the last equality, we use the identification (\ref{Ident}) with
$\vartheta$ given by (\ref{Concrete}).
Taking   the identifications (\ref{Double}) into account, 
we introduce a 
self-dual cone $\mathfrak{P}_{\mathrm{ext}}$ in $\h_{\mathrm{ext}}$ by 
\begin{align}
\mathfrak{P}_{\mathrm{ext}}=\mathscr{L}^2(L^2(\BbbR^d; dX))_+.
\end{align}

\begin{lemm}\label{DMultiOp}
Under the identifications (\ref{Double}),
we have the following:
\begin{itemize}
\item[{\rm (i)}]
$
V_n\otimes \one+\one \otimes V_n \unrhd 0
$ w.r.t. $\mathfrak{P}_{\mathrm{ext}}$.

\item[{\rm (ii)}]
$
f\otimes \one \pm \one \otimes f \unrhd 0
$
w.r.t. $\mathfrak{P}_{\mathrm{ext}}$ for each $f\in \mathfrak{A}_{\mathrm{e}}$.
\end{itemize}
\end{lemm}
{\it Proof.} 
We apply Ginibre's idea \cite{Ginibre}.

(i)
By the elementary fact
\begin{align}
\cos a+\cos b=2\cos \frac{a+b}{2} \cos \frac{a-b}{2}, \label{waseki}
\end{align} 
  we have
\begin{align}
V_n\otimes \one +\one \otimes V_n &=V_n(x_1)+V_n(x_2)\no
&=V_n\bigg(\frac{X_2-X_1}{\sqrt{2}}\bigg)+V_n\bigg(\frac{X_1+X_2}{\sqrt{2}}\bigg)\no
&=(2\pi)^{-d/2}\int_{\BbbR^d}
\hat{V}_n(p)
\Big\{
\cos\Big(
p\cdot\frac{X_2-X_1}{\sqrt{2}}
\Big)
+\cos\Big(
p\cdot\frac{X_2+X_1}{\sqrt{2}}
\Big)
\Big\}dp\no
&=(2\pi)^{-d/2}\int_{\BbbR^d}
2\underbrace{\hat{V}_n(p)}_{\ge 0}
\underbrace{\mathcal{L}
\Big[
\cos\Big(p\cdot \frac{X}{\sqrt{2}}\Big)
\Big]
\mathcal{R}\Big[
\cos\Big(p\cdot \frac{X}{\sqrt{2}}\Big)
\Big]}_{\unrhd 0\ \mathrm{by\  Proposition} \ \ref{GeneralPP}}
dp\no
&\unrhd 0\ \ \ \ \mathrm{w.r.t.} \ \ \  \mathfrak{P}_{\mathrm{ext}}.
\end{align}

(ii) By (\ref{waseki}) and 
\begin{align}
\cos a-\cos b=2 \sin \frac{b+a}{2} \sin\frac{b-a}{2}, \label{waseki2}
\end{align}
we have
\begin{align}
f\otimes \one +\one \otimes f&=(2\pi)^{-d/2}\int_{\BbbR^d} 2
\underbrace{\hat{f}(p)}_{\ge 0}
\underbrace{\mathcal{L}
\Big[
\cos\Big(p\cdot \frac{X}{\sqrt{2}}
\Big)
\Big]
\mathcal{R}\Big[
\cos\Big(p\cdot \frac{X}{\sqrt{2}}\Big)
\Big]}_{\unrhd 0\ \mathrm{by\  Proposition} \ \ref{GeneralPP}}dp\unrhd 0,\\
f\otimes \one -\one \otimes f&=(2\pi)^{-d/2}\int_{\BbbR^d} 2 
\underbrace{
\hat{f}(p)}_{\ge 0}
\underbrace{\mathcal{L}
\Big[
\sin\Big(p\cdot \frac{X}{\sqrt{2}}\Big)
\Big]
\mathcal{R}\Big[
\sin\Big(p\cdot \frac{X}{\sqrt{2}}\Big)
\Big]}_{\unrhd 0\ \mathrm{by\  Proposition} \ \ref{GeneralPP}}dp\unrhd 0
\end{align}
 w.r.t. $ \mathfrak{P}_{\mathrm{ext}}$. $\Box$

\begin{Thm}\label{DoublePP}
Assume that $V$ is regular. We have
$
e^{-\beta \mathbb{H}_n} \unrhd 0
$
w.r.t. $\mathfrak{P}_{\mathrm{ext}}$ for all $\beta \ge 0$.
\end{Thm}
{\it Proof.}
We set $\mathbb{H}_n=\mathbb{H}_0-\mathbb{V}_n$, where $\mathbb{H}_0=
(-\Delta_x)\otimes \one +\one \otimes(- \Delta_x)
$ and $\mathbb{V}_n
=V_n\otimes \one +\one \otimes V_n
$.
Note  that, by Lemma \ref{DMultiOp}, we know that $\mathbb{V}_n \unrhd
0$
w.r.t. $\mathfrak{P}_{\mathrm{ext}}$.
By (\ref{Delta}) and the identifications (\ref{Double}), we have
\begin{align}
\mathbb{H}_0=-\Delta_{X_1}-\Delta_{X_2}
=\mathcal{L}\Big(
-\Delta_X
\Big)
+
\mathcal{R}\Big(
-\Delta_X
\Big).
\end{align}
Thus,  by Proposition \ref{GeneralPP},
\begin{align}
e^{-\beta \mathbb{H}_0}=\mathcal{L}
\big[
e^{\beta \Delta_X}
\big]
\mathcal{R}
\big[
e^{\beta \Delta_X}
\big]
\unrhd 0 \label{LapD}
\end{align}
w.r.t. $\mathfrak{P}_{\mathrm{ext}}$. 
(Remark that, because $e^{\beta \Delta_x}$ is bounded, the RHS of (\ref{LapD}) 
is bounded as well.)
Now, we can apply Theorem \ref{StandPP}
and conclude the theorem. $\Box$

\begin{lemm}\label{Diff+-}
Let $f\in \mathfrak{A}_{\mathrm{e}}$.
Under the identifications (\ref{Double}), we have the following:
\begin{itemize}
\item[{\rm (i)}]  $
f(-i \nabla_x)\otimes \one +\one \otimes f(-i \nabla_x) \unrhd 0
$ w.r.t. $ \mathfrak{P}_{\mathrm{ext}}$. 
\item[{\rm (ii)}]
$
f(-i \nabla_x)\otimes \one -\one \otimes f(-i \nabla_x) \unlhd 0
$ w.r.t. $ \mathfrak{P}_{\mathrm{ext}}$. 
\end{itemize}
\end{lemm}
{\it Proof.} 
Note  that 
\begin{align}
\vartheta (-i \nabla_X) \vartheta^{-1}=+ i \nabla_X. \label{Conjugate}
\end{align}
(i) 
Since $f(-x)=f(x)$, we have 
\begin{align}
f(x)=(2\pi)^{-d/2} \int_{\BbbR^d} \hat{f}(p) \cos(p\cdot x)dp.\label{EvenF}
\end{align}
By (\ref{Coordinate2}),    (\ref{waseki}) and (\ref{EvenF}), 
\begin{align}
&f(-i \nabla_x)\otimes \one +\one \otimes f(-i \nabla_x) \no
&=f(-i \nabla_{x_1})+f(-i \nabla_{x_2})\no
&\underset{(\ref{Coordinate2})}{=}f\Big(
\frac{i}{\sqrt{2}} \nabla_{X_1}-\frac{i}{\sqrt{2}} \nabla_{X_2}
\Big)
+
f\Big(-\frac{i}{\sqrt{2}} \nabla_{X_1}-\frac{i}{\sqrt{2}} \nabla_{X_2}\Big)\no
&\underset{(\ref{EvenF})}{=}(2\pi)^{-d/2} \int_{\BbbR^d} \hat{f}(p)
\Bigg\{
\cos \Big[
p\cdot\Big(
\frac{i}{\sqrt{2}} \nabla_{X_1}-\frac{i}{\sqrt{2}} \nabla_{X_2}
\Big)
\Big]+\no
&\hspace{2cm}+
\cos \Big[
p\cdot\Big(
-\frac{i}{\sqrt{2}} \nabla_{X_1}-\frac{i}{\sqrt{2}} \nabla_{X_2}
\Big)
\Big]
\Bigg\} dp
\no
&\underset{(\ref{waseki})}{=} 2(2\pi)^{-d/2} \int_{\BbbR^d} \hat{f}(p) \cos\Big(
\frac{ -ip\cdot \nabla_{X_1}}{\sqrt{2}}
\Big)
\cos\Big(
\frac{-i p\cdot \nabla_{X_2}}{\sqrt{2}}
\Big) dp\no
&= 2(2\pi)^{-d/2}\int_{\BbbR^d} 
\underbrace{\hat{f}(p)}_{\ge 0}
\underbrace{
\mathcal{L}\Big[
\cos\Big(
\frac{-i p\cdot \nabla_X}{\sqrt{2}}
\Big)
\Big]
\mathcal{R}\Big[
\cos\Big(
\frac{-i p\cdot \nabla_X}{\sqrt{2}}
\Big)
\Big]
}_{\unrhd 0\ \mathrm{by\  Proposition} \ \ref{GeneralPP}}
dp\no
& \unrhd 0 \ \ \ \ \mbox{w.r.t. $ \mathfrak{P}_{\mathrm{ext}}$. }
\end{align}
This proves (i).
Similarly,  by (\ref{Coordinate2}), (\ref{waseki2}) and (\ref{EvenF}),
\begin{align}
&f(-i \nabla_x)\otimes \one -\one \otimes f(-i \nabla_x) \no
&\underset{(\ref{Coordinate2}) }{=}f\Big(
\frac{i}{\sqrt{2}} \nabla_{X_1}-\frac{i}{\sqrt{2}} \nabla_{X_2}
\Big)
-
f\Big(-\frac{i}{\sqrt{2}} \nabla_{X_1}-\frac{i}{\sqrt{2}} \nabla_{X_2}\Big)\no
&\underset{(\ref{EvenF})}{=}(2\pi)^{-d/2} \int_{\BbbR^d} \hat{f}(p)
\Bigg\{
\cos \Big[
p\cdot\Big(
\frac{i}{\sqrt{2}} \nabla_{X_1}-\frac{i}{\sqrt{2}} \nabla_{X_2}
\Big)
\Big]-\no
&\hspace{2cm}-
\cos \Big[
p\cdot\Big(
-\frac{i}{\sqrt{2}} \nabla_{X_1}-\frac{i}{\sqrt{2}} \nabla_{X_2}
\Big)
\Big]
\Bigg\}dp
\no
&\underset{(\ref{waseki2})}{=} 2 (2\pi)^{-d/2}\int_{\BbbR^d} \hat{f}(p) \sin\Big(
\frac{-i p\cdot \nabla_{X_1}}{\sqrt{2}}
\Big)
\sin\Big(
\frac{-i p\cdot \nabla_{X_2}}{\sqrt{2}
}
\Big) dp\no
&= 2(2\pi)^{-d/2}\int_{\BbbR^d} 
\hat{f}(p)
\mathcal{L}\Big[
\sin\Big(
\frac{-i p\cdot \nabla_X}{\sqrt{2}}
\Big)
\Big]
\mathcal{R}\Big[ 
\underbrace{
\vartheta 
\sin\Big(
\frac{-i p\cdot \nabla_X}{\sqrt{2}}
\Big)\vartheta^{-1}
}_{=- \sin\Big(
\frac{-i p\cdot \nabla_X}{\sqrt{2}}
\Big)\ \mathrm{by\ (\ref{Conjugate})}}
\Big]
dp\no
& \unlhd 0 \ \ \ \ \mbox{w.r.t. $ \mathfrak{P}_{\mathrm{ext}}$. }
\end{align} 
This proves (ii). $\Box$
\medskip\\

\subsection{Duhamel expansion}
Let $\Omega(x)=\pi^{-d/4} \mathrm{exp}(-|x|^2/2)\in L^2(\BbbR^d; dx)$
 and let 
$Z_{\beta, n}=\|e^{-\beta H_n} \Omega\|^2$.
We introduce a vector $\phi_{\beta, n}\in L^2(\BbbR^d; dx)$ by
\begin{align}
\phi_{\beta,n}=\frac{e^{-\beta H_n} \Omega}{\sqrt{Z_{\beta, n}}}. \label{PhiZ}
\end{align}

\begin{lemm}
$\displaystyle 
\la A\ra_n =\lim_{\beta \to \infty} \la \phi_{\beta, n}|A\phi_{\beta, n}\ra.
$
\end{lemm} 
{\it Proof.} 
By Proposition \ref{GSUni}, we have $\la
\Omega|\psi_n\ra>0$. 
Hence, we obtain 
\begin{align}
\psi_n=\mathrm{strong }\lim_{\beta \to \infty} \phi_{\beta, n}.
\end{align} 
Thus we are done. $\Box$

\begin{lemm}\label{DoublePositive}
Under the identifications (\ref{Double}), 
 we have
$\Omega\otimes \Omega \ge 0$ w.r.t. $\mathfrak{P}_{\mathrm{ext}}$.
\end{lemm} 
{\it Proof.}
By  (\ref{Coordinate}) and 
(\ref{Double}), 
\begin{align}
\Omega\otimes \Omega&=\pi^{-d/2}
 \mathrm{exp}\big\{-(X_1^2+X_2^2)/2\big\}=
\tilde{\Omega}\otimes \tilde{\Omega}=|\tilde{\Omega}\ra \la \tilde{\Omega}|, \label{VecDou}
\end{align} 
where $\tilde{\Omega}(X)=\pi^{-d/4} \exp(-|X|^2/2)\in L^2(\BbbR^d; dX)$.
The RHS of (\ref{VecDou}) $\ge 0$ w.r.t. $\mathfrak{P}_{\mathrm{ext}}$,
 because the projection $|\tilde{\Omega}\ra \la \tilde{\Omega}|$ is
 positive as a  linear operator on $L^2(\BbbR^d; dX)$.
$\Box$

\begin{Thm}\label{A+-}
Let $A\in \mathscr{B}(L^2(\BbbR^d; dx))$. 
\begin{itemize}
\item[{\rm (i)}] If $A\otimes \one -\one \otimes A \unrhd 0$
	     w.r.t. $\mathfrak{P}_{\mathrm{ext}}$, then 
$\la A\ra_n$ is monotonically increasing in $n$.
\item[{\rm (ii)}] If $A\otimes \one -\one \otimes A \unlhd 0$
	     w.r.t. $\mathfrak{P}_{\mathrm{ext}}$, then 
$\la A\ra_n$ is monotonically decreasing  in $n$.
\end{itemize} 
\end{Thm} 
{\it Proof.} Suppose that  $n\ge m$.  Note that 
\begin{align}
\la A\ra_n-\la A\ra_m
&=\lim_{\beta\to \infty}\frac{Z_{\beta, m}}{ Z_{\beta, n}}
 \mathscr{I}_{\beta},
\label{SaKey}
\end{align} 
where
\begin{align}
\mathscr{I}_{\beta}=
\frac{
\big\la e^{-\beta H_n}\Omega\big|Ae^{-\beta
 H_n}\Omega\big\ra}{Z_{ \beta, m}} 
-
\frac{
\big\la e^{-\beta H_m}\Omega\big|Ae^{-\beta
 H_m}
\Omega
\big\ra
 }
{
Z_{ \beta, m}
}
\frac{ Z_{ \beta, n}}{Z_{ \beta, m}}. \label{BasicCh}
\end{align} 

Let $\delta=V_n-V_m$.
By the Duhamel formula, 
\begin{align}
e^{-\beta H_n}=e^{-\beta(H_m-\delta)}
=\sum_{j\ge 0} \int_{\mathcal{T}_j(\beta)}
\delta(s_1)\cdots \delta(s_n)   e^{-\beta H_m} ds_1\cdots ds_n , \label{Duha}
\end{align} 
where $\delta(s)=e^{-sH_m} \delta e^{s H_m}$ and $\mathcal{T}_j(\beta)
=\{(s_1, \dots, s_j)\, |\, 0\le s_1\le \cdots \le s_j\le \beta\}
$.
The RHS of  (\ref{Duha}) converges in the operator norm topology.

For each $A\in \mathscr{B}(L^2(\BbbR^d; dx))$, we set
\begin{align}
\omega(A)=\la \phi_{\beta, m}|A\phi_{\beta, m}\ra.
\end{align}  

The following formula is useful:
\begin{lemm}
We have
\begin{align}
\mathscr{I}_{\beta}
=\sum_{i, j\ge 0}
\int_{\mathcal{T}_i(\beta)}\int_{\mathcal{T}_j(\beta)}
\Big\{
\omega\big(
X_i(\Bs)AY_j(\Bt)
\big) 
-\omega(A)
\omega\big(
 X_i(\Bs)Y_j(\Bt)\big)
\Big\}ds_1\cdots ds_i dt_1\cdots dt_j
, \label{Ipositive}
\end{align}
where 
$
X_i(\Bs)=\delta(s_i)\delta(s_{i-1})\cdots \delta(s_1) $ and $  
Y_j(\Bt)=\delta(t_1)\cdots \delta(t_{j-1})\delta(t_j).$
\end{lemm}
{\it Proof.} By (\ref{PhiZ}) and (\ref{Duha}), 
we have 
\begin{align}
&\frac{
\big\la e^{-\beta H_n}\Omega\big|Ae^{-\beta
 H_n}\Omega\big\ra}{Z_{ \beta, m}} \no
 =&\sum_{i, j\ge 0}
\int_{\mathcal{T}_i(\beta)}\int_{\mathcal{T}_j(\beta)}
Z_{\beta, m}^{-1}
\Big\la e^{-\beta H_m}\Omega\Big| X_i(\Bs) AY_j(\Bt) e^{-\beta H_m}\Omega\Big\ra ds_1\cdots ds_i dt_1\cdots dt_j\no
=& \sum_{i, j\ge 0} \int_{\mathcal{T}_i(\beta)}\int_{\mathcal{T}_j(\beta)}
\omega \big(X_i(\Bs) AY_j(\Bt) \big) ds_1\cdots ds_i dt_1\cdots dt_j.
 \label{IntFormA}
\end{align}
As for the term $Z_{\beta, n}/Z_{\beta, m}$, we have, by (\ref{IntFormA}),
\begin{align}
\frac{Z_{\beta, n}}{Z_{\beta, m}}&=\frac{
\big\la e^{-\beta H_n}\Omega\big|\one e^{-\beta
 H_n}\Omega\big\ra}{Z_{ \beta, m}} \no
 &= \sum_{i, j\ge 0} \int_{\mathcal{T}_i(\beta)}\int_{\mathcal{T}_j(\beta)}
\omega \big(X_i(\Bs) Y_j(\Bt) \big) ds_1\cdots ds_i dt_1\cdots dt_j.
\end{align}
Inserting these formulas into (\ref{BasicCh}), we obtain the desired identity. $\Box$
\medskip\\

Thus, to prove the theorem, it suffices to prove the following proposition.

\begin{Prop}\label{PPPPP}
Let $A\in \mathscr{B}(L^2(\BbbR^d;  dx))$. 
 
\begin{itemize}
\item[{\rm (i)}] If $A\otimes \one -\one \otimes A \unrhd 0$
	     w.r.t. $\mathfrak{P}_{\mathrm{ext}}$, then
we have
\begin{align}
\omega\big( X_i(\Bs)AY_j(\Bt)\big)
-
\omega\big( X_i(\Bs)Y_j(\Bt)\big)\omega(  A)\ge 0
\end{align}
for all $\Bs\in \mathcal{T}_i(\beta)$ and $\Bt\in
	     \mathcal{T}_j(\beta)$.
\item[{\rm (ii)}] If $A\otimes \one -\one \otimes A \unlhd 0$
	     w.r.t. $\mathfrak{P}_{\mathrm{ext}}$, then
we have
\begin{align}
\omega\big( X_i(\Bs)AY_j(\Bt)\big)-
\omega\big(X_i(\Bs)Y_j(\Bt)\big)\omega(  A)\le 0
\end{align}
for all $\Bs\in \mathcal{T}_i(\beta)$ and $\Bt\in
	     \mathcal{T}_j(\beta)$.
\end{itemize} 
\end{Prop} 
{\it Proof.} (i)
For each $B\in \mathscr{B}(L^2(\BbbR^d; dx))$, we set
\begin{align}
B_{\pm}=B\otimes \one\pm \one \otimes B.
\end{align} 
By (\ref{waseki}), 
\begin{align}
\delta_+=2 (2\pi)^{-d/2}\int_{\BbbR^d}
 \underbrace{(\hat{V}_n(p)-\hat{V}_m(p))}_{\ge 0}
\underbrace{
\mathcal{L}\Big[
\cos\Big(\frac{p\cdot X}{\sqrt{2}}\Big)
\Big]
\mathcal{R}\Big[
\cos\Big(\frac{p\cdot X}{\sqrt{2}}\Big)
\Big]
}_{\unrhd 0}
dp
\unrhd 0\ \ \ \mbox{w.r.t. $\mathfrak{P}_{\mathrm{ext}}$}.\label{DeltaP}
\end{align} 
Similarly,  $\delta_-\unrhd 0$ w.r.t. $\mathfrak{P}_{\mathrm{ext}}$.
In addition, 
$
A_{-} \unrhd 0$ w.r.t. $\mathfrak{P}_{\mathrm{ext}}$ by the assumption.

We define
\begin{align}
X_{\pm}(\Bs)=\Bigg[
\prod^{{i}\atop{\longleftarrow}}_{\alpha=1} \delta(s_{\alpha})
\Bigg]\otimes \one 
\pm
\one 
\otimes 
\Bigg[
\prod^{{i}\atop{\longleftarrow}}_{\alpha=1} \delta(s_{\alpha})
\Bigg],
\end{align} 
where $\displaystyle 
\prod^{{i}\atop{\longleftarrow}}_{\alpha=1}B_{\alpha}=B_iB_{i-1}\cdots
B_2 B_1
$, an ordered product. Let 
\begin{align}
\delta_{\pm}[s]
=e^{-s \mathbb{H}_m} \delta_{\pm} e^{s \mathbb{H}_m}.
\end{align}
Since $\delta\otimes \one =\frac{1}{2}(\delta_++\delta_-)$ and $\one
\otimes \delta=\frac{1}{2}(\delta_+-\delta_-)$, we obtain
\begin{align}
X_{\pm}(\Bs)=
2^{-i} \prod^{{i}\atop{\longleftarrow}}_{\alpha=1}
\Bigg\{
\delta_+[s_{\alpha}]+\delta_-[s_{\alpha}]
\Bigg\}
\pm
2^{-i} \prod^{{i}\atop{\longleftarrow}}_{\alpha=1}
\Bigg\{
\delta_+[s_{\alpha}]-\delta_-[s_{\alpha}]
\Bigg\}. \label{SumDiff}
\end{align}

For each ${\boldsymbol \vepsilon}=\{\vepsilon_1, \dots, \vepsilon_i\}
\in \{+, -\}^i$, we define 
\begin{align}
{\boldsymbol \delta}_{{\boldsymbol \vepsilon}}[\Bs]
=
 \prod^{{i}\atop{\longleftarrow}}_{\alpha=1} \delta_{\vepsilon_{\alpha}}[s_{\alpha}].
\end{align}
In terms of this notation, 
\begin{align}
\prod^{{i}\atop{\longleftarrow}}_{\alpha=1}
\Bigg\{
\delta_+[s_{\alpha}]+\delta_-[s_{\alpha}]
\Bigg\}
&=\sum_{{\boldsymbol {\vepsilon}} \in \{+, -\}^i} {\boldsymbol
 \delta}_{{\boldsymbol \vepsilon}}[\Bs],\\
\prod^{{i}\atop{\longleftarrow}}_{\alpha=1}
\Bigg\{
\delta_+[s_{\alpha}]-\delta_-[s_{\alpha}]
\Bigg\}
&=\sum_{{\boldsymbol {\vepsilon}} \in \{+, -\}^i} 
\sigma(
{\boldsymbol {\vepsilon}}
)
{\boldsymbol
 \delta}_{{\boldsymbol \vepsilon}}[\Bs],
\end{align} 
where $\sigma({\boldsymbol \vepsilon})=(\vepsilon_1 1)(\vepsilon_2 1)\cdots
(\vepsilon_i 1)
=+1
$ if the number of $\vepsilon_{\alpha}=-$ is even,
 $\sigma(
{\boldsymbol \vepsilon }
)=-1$  if the number of $\vepsilon_{\alpha}=-$ is odd.
Thus, we have
\begin{align}
X_{+}(\Bs)
=2^{-(i-1)}  \sum_{\sigma({\boldsymbol  \vepsilon})=+1} {\boldsymbol
 \delta}_{{\boldsymbol \vepsilon}}[\Bs],
\ \ \ \ 
X_{-}(\Bs)
=2^{-(i-1)}  \sum_{\sigma({\boldsymbol  \vepsilon})=-1} {\boldsymbol
 \delta}_{{\boldsymbol \vepsilon}}[\Bs]. \label{XEx}
\end{align} 
Because, for each $\Bs\in \mathcal{T}_i(\beta)$,
\begin{align}
e^{-\beta \mathbb{H}_m}{\boldsymbol \delta}_{{\boldsymbol
 \vepsilon}}[\Bs]
=
\underbrace{
e^{-(\beta-s_i) \mathbb{H}_m}
}_{\unrhd 0}
\underbrace{
 \delta_{\vepsilon_i}
}_{\unrhd 0}
\underbrace{
e^{-(s_i-s_{i-1}) \mathbb{H}_m}
}_{\unrhd 0}
\cdots
\underbrace{
e^{-(s_2-s_1) \mathbb{H}_m} 
}_{\unrhd 0}
\underbrace{
\delta_{\vepsilon_1}
}_{\unrhd 0}
\underbrace{
 e^{-s_1\mathbb{H}_m}
}_{\unrhd 0}
\unrhd 0
\end{align} 
w.r.t. $\mathfrak{P}_{\mathrm{ext}}$, we conclude that 
$e^{-\beta \mathbb{H}_m}X_{\pm}(\Bs) \unrhd 0$ w.r.t. $\mathfrak{P}_{\mathrm{ext}}$
by (\ref{XEx}).
Similarly, we can prove that 
$
Y_{\pm}(\Bt) e^{-\beta \mathbb{H}_m}
\unrhd 0
$ w.r.t. $\mathfrak{P}_{\mathrm{ext}}$.

Because 
\begin{align}
\underbrace{e^{-\beta \mathbb{H}_m}X_+(\Bs)}_{\unrhd 0}
\underbrace{A_-}_{\unrhd 0}
\underbrace{Y_-(\Bt) e^{-\beta \mathbb{H}_m}}_{\unrhd 0}
\unrhd 0
\end{align}
w.r.t. $\mathfrak{P}_{\mathrm{ext}}$, we have, by Lemma \ref{DoublePositive}, 
\begin{align}
&\bigg\la \phi_{\beta, m}\otimes \phi_{\beta, m}\bigg| 
X_+(\Bs)
A_-
Y_-(\Bt)
\phi_{\beta, m}\otimes \phi_{\beta, m}
\bigg\ra \no
&=Z_{\beta, n}^{-2}
\bigg\la 
\underbrace{\Omega\otimes \Omega}_{\ge 0}
\bigg|
\underbrace{
e^{-\beta \mathbb{H}_m} 
X_+(\Bs)
A_-
Y_-(\Bt)
e^{-\beta \mathbb{H}_m}
}_{\unrhd 0}
\underbrace{
\Omega\otimes \Omega
}_{\ge 0}
\bigg\ra \ge 0,
\end{align} 
implying  that 
\begin{align}
&\omega\big( X_i(\Bs)AY_j(\Bt)\big)
-
\omega\big( 
 X_i (\Bs)Y_j(\Bt)\big)
\omega( A)\no
&\ \ 
+\omega\big(  AY_j(\Bt)\big)
\omega\big( X_i(\Bs)\big)
-
\omega\big( Y_j(\Bt)\big) \omega\big( X_i(\Bs)A\big)\ge 0. \label{+--}
\end{align}
On the other hand, we have
$
e^{-\beta \mathbb{H}_m}
X_-(\Bs)
A_-
Y_+(\Bt)
e^{-\beta \mathbb{H}_m}
\unrhd 0
$
w.r.t. $\mathfrak{P}_{\mathrm{ext}}$, which implies
\begin{align}
&\omega\big( X_i(\Bs)AY_j(\Bt) \big)
-
\omega\big(
 X_i(\Bs)Y_j(\Bt)
\big)
\omega(  A)\no
&\ \ \ - 
\omega\big(  AY_j(\Bt)\big)
\omega\big( 
 X_i(\Bs)
\big)
+
\omega\big( Y_j(\Bt)\big)
\omega\big( X_i(\Bs)A\big)\ge 0.\label{--+}
\end{align}
Combining (\ref{+--}) and (\ref{--+}), we obtain the desired result. 
We can prove (ii) similarly. $\Box$

\begin{flushleft}
{\it 
Proof of Theorem \ref{MonoEx}
}
\end{flushleft} 
By Lemma \ref{Diff+-} and Theorem \ref{A+-}, we conclude Theorem \ref{MonoEx}. $\Box$

\section{Proof of Theorem \ref{GriSec}}\label{6}
\setcounter{equation}{0}
We begin with the following proposition.

\begin{Prop}\label{DoubleMono}
If $n>m$, then $e^{-\beta \mathbb{H}_n} \unrhd e^{-\beta \mathbb{H}_m} \unrhd 0$
w.r.t. $\mathfrak{P}_{\mathrm{ext}}$ for all $\beta \ge 0$.

\end{Prop}
{\it Proof.} 
By (\ref{DeltaP}), we already know that
$\delta_+=\mathbb{V}_n-\mathbb{V}_m
\unrhd 0$ 
 w.r.t. $\mathfrak{P}_{\mathrm{ext}}$.
Because 
$
\mathbb{H}_n=\mathbb{H}_m-\delta_+ 
$,
 we conclude the assertion
by using Theorem \ref{Mono}. $\Box$
\medskip\\

Let 
\begin{align}
\mathbb{H}=H\otimes \one +\one \otimes H.
\end{align}

\begin{Thm}\label{DoublePP}
$e^{-\beta \mathbb{H}}\unrhd 0$ w.r.t. $\mathfrak{P}_{\mathrm{ext}}$ for
 all $\beta \ge 0$.
\end{Thm} 
{\it Proof.} 
By Proposition \ref{DoubleMono}, we know that  $e^{-\beta \mathbb{H}_n} \unrhd e^{-\beta \mathbb{H}_m} \unrhd 0$
w.r.t. $\mathfrak{P}_{\mathrm{ext}}$ for all $\beta \ge 0$, provided
that $n>m$. Since $e^{-\beta \mathbb{H}_n}$ strongly converges to
$e^{-\beta \mathbb{H}}$ by the assumption {\bf (B)}, we obtain
$
e^{-\beta \mathbb{H}} \unrhd e^{-\beta \mathbb{H}_m}\unrhd 0
$ w.r.t. $\mathfrak{P}_{\mathrm{ext}}$ for all $\beta \ge 0$ by
Proposition \ref{Basic}. $\Box$

\begin{coro}\label{PPDoubleGs}
Let $\psi$ be the unique ground state of $H$. Under the
 identifications (\ref{Double}), 
$\psi\otimes \psi\ge 0$ w.r.t. $\mathfrak{P}_{\mathrm{ext}}$.
\end{coro}
{\it Proof.} 
Let $\Psi=\psi\otimes \psi$. Since the ground state of $H$
is unique, $\Psi$ is the unique ground state of $\mathbb{H}$.
Thus, by Proposition \ref{GSPosi} and Theorem  \ref{DoublePP}, we conclude the assertion. $\Box$

\begin{Thm}\label{Combi+-}
Let $A, B\in \mathscr{B}(L^2(\BbbR^d; dx))$.
Under the identifications (\ref{Double}), we have the following:
\begin{itemize}
\item[{\rm (i)}]
If $A\otimes \one -\one \otimes A\unrhd 0$ and 
$B\otimes \one -\one \otimes B\unrhd 0$
	     w.r.t. $\mathfrak{P}_{\mathrm{ext}}$, then
$\la AB\ra-\la A\ra\la B\ra\ge 0$.
\item[{\rm (ii)}]
If $A\otimes \one -\one \otimes A\unlhd 0$ and 
$B\otimes \one -\one \otimes B\unlhd 0$
	     w.r.t. $\mathfrak{P}_{\mathrm{ext}}$, then
$\la AB\ra-\la A\ra\la B\ra\ge 0$.
\item[{\rm (iii)}]
If $A\otimes \one -\one \otimes A\unlhd 0$ and 
$B\otimes \one -\one \otimes B\unrhd 0$
	     w.r.t. $\mathfrak{P}_{\mathrm{ext}}$, then
$\la AB\ra-\la A\ra\la B\ra\le 0$.
\end{itemize} 
\end{Thm} 
{\it Proof.}
(i)
By Corollary \ref{PPDoubleGs}, 
\begin{align}
2(\la AB\ra-\la A\ra\la B\ra)
=\Big\la 
\underbrace{
\psi\otimes \psi
}_{\ge 0}
\Big|
\underbrace{
(
A\otimes \one -\one \otimes A
)
}_{\unrhd 0}
\underbrace{
(
B\otimes \one -\one \otimes B
)
}_{\unrhd 0}
\underbrace{
\psi\otimes \psi
}_{\ge 0}
\Big\ra\ge 0.
\end{align} 
Thus, we obtain (i). We can prove (ii) and (iii) similarly. $\Box$
\begin{flushleft}
{\it Proof of Theorem \ref{GriSec}}
\end{flushleft} 
By Lemmas \ref{DMultiOp}, \ref{Diff+-} and Theorem \ref{Combi+-}, we conclude Theorem \ref{GriSec}. $\Box$

\section{Proof of Theorem \ref{MonoG}}\label{7}
\setcounter{equation}{0}

Let $V_{ n}^{(1)}$ (resp.,  $V_{ n}^{(2)}$) be an approximating sequence of $V^{(1)}$
(resp.,  $V^{(2)}$) in  condition {\bf (B)}. Let 
\begin{align}
H_{n}^{(1)}=-\Delta_x-V_{n}^{(1)},\ \ \ H_n^{(2)}=-\Delta_x-V_{n}^{(2)}.
\end{align} 
Then,
\begin{align}
H_n^{(1)}=H_{n}^{(2)}-W_n,\ \ \ W_n=V_{n}^{(1)}-V_{ n}^{(2)}. \label{DiffH}
\end{align} 
As previously, we study the extended Hamiltonian
\begin{align}
\mathbb{H}_n^{(1)}=H_n^{(1)}\otimes \one +\one \otimes H_n^{(1)},\ \ \ \ 
\mathbb{H}_n^{(2)}=H_n^{(2)}\otimes \one +\one \otimes H_n^{(2)}.
\end{align} 
By (\ref{DiffH}), 
\begin{align}
\mathbb{H}_n^{(1)}=\mathbb{H}_n^{(2)}-\mathbb{W}_n,\ \ \ \ \mathbb{W}_n=
W_n\otimes \one +\one \otimes W_n.
\end{align} 
\begin{lemm}
$
\mathbb{W}_n\unrhd 0
$ w.r.t. $\mathfrak{P}_{\mathrm{ext}}$.
\end{lemm} 
{\it Proof.} In a similar manner as in the proof of Lemma \ref{DMultiOp} (i), we
see that
\begin{align}
\mathbb{W}_n
=2(2\pi)^{-d/2} \int_{\BbbR^d}
 \underbrace{\big(\hat{V}_n^{(1)}(k)-\hat{V}_n^{(2)}(k)\big)}_{\ge 0}
\underbrace{
\mathcal{L}
\Big[
\cos\Big(\frac{k\cdot X}{\sqrt{2}}\Big)
\Big]
\mathcal{R} 
\Big[
\cos\Big(\frac{k\cdot X}{\sqrt{2}}\Big)
\Big]
}_{\unrhd 0}
dk
\unrhd 0
\end{align}
 w.r.t. $\mathfrak{P}_{\mathrm{ext}}$. $\Box$

\begin{Thm}\label{H1H2Comp}
Let $A\in \mathscr{B}(L^2(\BbbR^d; dx))$. 
\begin{itemize}
\item[{\rm (i)}] If $A\otimes \one -\one \otimes A\unrhd 0$
	     w.r.t. $\mathfrak{P}_{\mathrm{ext}}$, then 
$\la A\ra^{(1)} \ge \la A\ra^{(2)}$.
\item[{\rm (ii)}] If $A\otimes \one -\one \otimes A\unlhd 0$
	     w.r.t. $\mathfrak{P}_{\mathrm{ext}}$, then 
$\la A\ra^{(1)} \le \la A\ra^{(2)}$.
\end{itemize} 
\end{Thm} 
{\it Proof.} The proof of this theorem is similar to that of Theorem \ref{A+-}.
Hence, we provide only  a sketch of the proof.
Let $\psi_n^{(1)}$ (resp., $\psi_n^{(2)}$) be the unique ground state of
$H_n^{(1)}$
 (resp.,  $H_n^{(2)}$).
For each $A\in \mathscr{B}(L^2(\BbbR^d; dx))$, we set
\begin{align}
\la A\ra_n^{(1)}=\big\la \psi_n^{(1)}\big|A\psi_n^{(1)}\big\ra,\ \ \ 
\la A\ra_n^{(2)}=\big\la \psi_n^{(2)}\big|A\psi_n^{(2)}\big\ra.
\end{align} 
Corresponding to (\ref{SaKey}), we obtain
\begin{align}
\la A\ra_n^{(1)}-\la A\ra_n^{(2)}
=\lim_{\beta\to \infty} \frac{Z_{\beta}^{(2)}}{Z_{\beta}^{(1)}} \mathscr{J}_{\beta},
\end{align} 
where $Z_{\beta}^{(j)}=\big\|e^{-\beta H_n^{(j)}} \Omega\big\|^2\ (j=1,2)$
and  
\begin{align}
\mathscr{J}_{\beta}
=\frac{
\Big\la e^{-\beta H_n^{(1)}} \Omega\Big| Ae^{-\beta H_n^{(1)}} \Omega \Big\ra
}
{Z_{\beta}^{(2)}}
-
\frac{
\Big\la e^{-\beta H_n^{(2)}} \Omega\Big| Ae^{-\beta H_n^{(2)}} \Omega \Big\ra
}
{Z_{\beta}^{(2)}}
\frac{Z_{\beta}^{(1)}}{Z_{\beta}^{(2)}}.
\end{align} 
Since $\displaystyle
\la A \ra^{(\alpha)}=\lim_{n\to \infty} \la A \ra_n^{(\alpha)}
$ for each $\alpha=1,2$, it suffices to prove that
$\mathscr{J}_{\beta}\ge 0$ for all $\beta >0$.

Let $\displaystyle 
\phi_n^{(2)}=e^{-\beta H_n^{(2)}}\Omega\Big/\sqrt{Z_{\beta}^{(2)}}
$. We set
\begin{align}
\tomega(A)=\Big\la \phi_n^{(2)}\Big|A\phi_n^{(2)}\Big\ra,\ \ A\in
 \mathscr{B}(L^2(\BbbR^d; dx)).
\end{align} 
By the Duhamel formula, we obtain
\begin{align}
\mathscr{J}_{\beta}
=\sum_{i, j\ge 0} \int_{\mathcal{T}_i(\beta)}
 \int_{\mathcal{T}_j(\beta)}
\Big\{
\tomega\Big(
\mathcal{X}_i(\Bs) A\mathcal{Y}_j(\Bt)
\Big)
-
\tomega(A)
\tomega\Big(
\mathcal{X}_i(\Bs) \mathcal{Y}_j(\Bt)
\Big)
\Big\}ds_1\cdots ds_i dt_1\cdots dt_j, \label{JB}
\end{align} 
where  $
\mathcal{X}_i(\Bs)=W_n(s_i)W_n(s_{i-1})\cdots W_n(s_1)$ and $
\mathcal{Y}_j(\Bt)=W_n(t_1)\cdots W_n(t_{j-1})W_n(t_j).$
By Proposition \ref{WWW} below, the RHS of (\ref{JB}) is positive. $\Box$
\medskip\\

\begin{Prop}\label{WWW}
Let $A\in \mathscr{B}(L^2(\BbbR^d;  dx))$. 
 
\begin{itemize}
\item[{\rm (i)}] If $A\otimes \one -\one \otimes A \unrhd 0$
	     w.r.t. $\mathfrak{P}_{\mathrm{ext}}$, then 
we have
\begin{align}
\tomega\big( \mathcal{X}_i(\Bs)A \mathcal{Y}_j(\Bt)\big)
-
\tomega\big( \mathcal{X}_i(\Bs) \mathcal{Y}_j(\Bt)\big)\tomega(  A)\ge 0
\end{align}
for all $\Bs\in \mathcal{T}_i(\beta)$ and $\Bt\in
	     \mathcal{T}_j(\beta)$.
\item[{\rm (ii)}] If $A\otimes \one -\one \otimes A \unlhd 0$
	     w.r.t. $\mathfrak{P}_{\mathrm{ext}}$, then 
we have
\begin{align}
\tomega\big( \mathcal{X}_i(\Bs)A \mathcal{Y}_j(\mathbf{t})\big)-
\tomega\big(\mathcal{X}_i(\Bs) \mathcal{Y}_j(\Bt)\big)\tomega(  A)\le 0
\end{align}
for all $\Bs\in \mathcal{T}_i(\beta)$ and $\Bt\in
	     \mathcal{T}_j(\beta)$.
\end{itemize} 
\end{Prop} 
{\it Proof.} We can prove  Proposition \ref{WWW}  in a manner similar
to that  in the proof of Proposition \ref{PPPPP}. $\Box$
\begin{flushleft}
{\it Proof of Theorem \ref{MonoG}}
\end{flushleft} 
By Lemmas \ref{DMultiOp}, \ref{Diff+-} and Theorem \ref{H1H2Comp}, we conclude Theorem
\ref{MonoG}. $\Box$

\section{Proof of Theorems \ref{PPGS1}, \ref{PPGS2} and \ref{PPGS3}} \label{NewApp}
\setcounter{equation}{0}
\subsection{Proof of Theorem \ref{PPGS1}}
For each $f\in \mathfrak{A}$ and  $a\in \BbbR^d$, we set
\begin{align}
\mathscr{C}_a^{+}[f]&= f+\frac{1}{2}\{
f(\cdot -a)+f(\cdot +a)
\},\\
\mathscr{C}_a^{-}[f]&= f-\frac{1}{2}\{
f(\cdot -a)+f(\cdot +a)
\}.
\end{align}

\begin{Prop}\label{Cafe}
$\mathscr{C}_a^{\pm}$ maps $\mathfrak{A}$ into $\mathfrak{A}$.
\end{Prop}
{\it Proof.}
Let $\hat{\mathscr{C}}^{\pm}_a[f]$ be the Fourier transform of $
\mathscr{C}_a^{\pm}[f]
$. We have 
\begin{align}
\hat{\mathscr{C}}_a^{\pm}[f](p) &=\{1\pm\cos( p\cdot a) \}\hat{f}(p) \ge 0.
\end{align}
Thus we are done. $\Box$

\begin{flushleft}
{\it Proof of Theorem \ref{PPGS1}}
\end{flushleft}

(i) 
Choose $a\in \mathcal{C}(V)$ arbitrarily.
By Theorem \ref{FirstInq} (i) and Proposition \ref{Cafe}, we have
$\la \mathscr{C}^-_a[f]\ra \ge 0$. By a limiting argument,\footnote{
To be precise, 
take $f\in C_0^{\infty}(\BbbR^d)$ with $\|f\|_{L^2}=1$.
Set $f_{\vepsilon}(x)=\vepsilon^{-d/2}f(x/\vepsilon) $.
 Then we have, by the dominated convergence theorem,
\begin{align}
\int_{\BbbR^d} \psi(x)^2 f_{\vepsilon}(x)dx\to \psi(0)^2 
\end{align}
as $\vepsilon\to +0$. 
Thus, $\la \mathscr{C}_a^-[f]\ra\ge 0$ implies (\ref{CaPlus}).
} we have
\begin{align}
\psi(0)^2-\frac{1}{2}\{\psi(a)^2+\psi(-a)^2\} \ge 0. \label{CaPlus}
\end{align}
Because $\psi(-a)=\psi(a)$, we obtain the desired result.

(ii) 
Let $p\in \hat{\mathcal{C}}(V)$.
By Theorem \ref{FirstInq} (ii) and Proposition \ref{Cafe}, we have
$\la \mathscr{C}^-_p[f](-i \nabla_x)\ra \ge 0$. 
Since $\la f(-i\nabla_x)\ra=\la\hat{\psi}|f\hat{\psi}\ra$, we have 
$
\la \hat{\psi}|\mathscr{C}_p^-[f]\hat{\psi}\ra\ge 0
$.
By a limiting argument, we have
\begin{align}
\hat{\psi}(0)^2-\frac{1}{2}\{\hat{\psi}(p)^2+\hat{\psi}(-p)^2\} \ge 0.
\end{align}
Because $\hat{\psi}(-p)=\hat{\psi}(p)$, we conclude the assertion. $\Box$

\subsection{Proof of Theorems \ref{PPGS2} and \ref{PPGS3}}

\begin{Prop}\label{Caf}
$\mathscr{C}_a^{\pm}$ maps $\mathfrak{A}_{\mathrm{e}}$ into $\mathfrak{A}_{\mathrm{e}}$.
\end{Prop}
{\it Proof.}
It is easy to check that $
\mathscr{C}_a^{\pm}[f](-x)=\mathscr{C}_a^{\pm}[f](x)
$.
Thus, the assertion follows from Proposition \ref{Cafe}. $\Box$

\begin{flushleft}
{\it Proof of Theorem \ref{PPGS2}}
\end{flushleft}
Let $a\in \mathcal{C}(V^{(1)}) \cap \mathcal{C}(V^{(2)})$.
By Theorem \ref{MonoG} and Proposition \ref{Caf}, we have 
$
 \big\la \mathscr{C}_a^{\pm}[f] \big\ra^{(1)}
 \ge 
 \big\la \mathscr{C}_a^{\pm}[f] \big\ra^{(2)}
$.
By a limiting argument, we obtain that 
\begin{align}
 \psi^{(1)}(0)^2\pm\frac{1}{2} \{
 \psi^{(1)}(a)^2+\psi^{(1)}(-a)^2
 \}
 \ge 
 \psi^{(2)}(0)^2\pm\frac{1}{2} \{
 \psi^{(2)}(a)^2+\psi^{(2)}(-a)^2
 \}.
\end{align}
Because $\psi^{(j)}(-x)=\psi^{(j)}(x)$,   we have
\begin{align}
\psi^{(1)}(0)^2\pm\psi^{(1)}(a)^2&\ge \psi^{(2)}(0)^2\pm\psi^{(2)}(a)^2.
\end{align}
Thus we are done. $\Box$

\begin{flushleft}
{\it Proof of Theorem \ref{PPGS3}}
\end{flushleft}
Choose $p\in \hat{\mathcal{C}}(V^{(1)}) \cap \hat{\mathcal{C}}(V^{(2)})$ arbitrarily.
By Theorem \ref{MonoG} and Proposition \ref{Caf}, we have 
$
 \big\la \mathscr{C}_p^{\pm}[f] (-i\nabla_x)\big\ra^{(1)}
 \le 
 \big\la \mathscr{C}_p^{\pm}[f] (-i\nabla_x)\big\ra^{(2)}
$.
Because 
$
\la f(-i \nabla_x)\ra^{(j)}=\la \hat{\psi}^{(j)}|f\hat{\psi}^{(j)}\ra,\ j=1,2
$, we have 
$
\la \hat{\psi}^{(1)}|\mathscr{C}_p^{\pm}[f] \hat{\psi}^{(1)}\ra\le 
\la \hat{\psi}^{(2)}|\mathscr{C}_p^{\pm}[f] \hat{\psi}^{(2)}\ra
$.
By a limiting argument, we obtain that 
\begin{align}
 \hat{\psi}^{(1)}(0)^2\pm\frac{1}{2} \{
 \hat{\psi}^{(1)}(p)^2+\hat{\psi}^{(1)}(-p)^2
 \}
 \le 
 \hat{\psi}^{(2)}(0)^2\pm\frac{1}{2} \{
 \hat{\psi}^{(2)}(p)^2+\hat{\psi}^{(2)}(-p)^2
 \}.
\end{align}
Because $\hat{\psi}^{(j)}(-p)=\hat{\psi}^{(j)}(p)$,   we have
\begin{align}
\hat{\psi}^{(1)}(0)^2\pm\hat{\psi}^{(1)}(p)^2\le \hat{\psi}^{(2)}(0)^2\pm\hat{\psi}^{(2)}(p)^2.
\end{align}
This completes the proof. $\Box$

\section{Proof of Theorems \ref{App1}, \ref{App2} and \ref{App3}}\label{8}
\setcounter{equation}{0}
\subsection{Proof of Theorem \ref{App1}}

(i) 
By Theorem \ref{FirstInq} (i), 
\begin{align}
\la f\ra=(2\pi)^{-d/2} \int_{\BbbR} dp \hat{f}(p) \la \psi|e^{ip\cdot
 x}\psi\ra
=\int_{\BbbR^d} dp\hat{f}(p) \hat{\varrho}(p)>0
\end{align} 
for all $f\in \mathfrak{A}\cap L^1(\BbbR^d; dx)$ with $f\neq 0$.
 Thus, we conclude (i).

(ii) Since $V(-x)=V(x)$ a.e. $x$ by the assumption (ii) of  {\bf (B)}, we
know that $\psi(-x)=\psi(x)$ a.e. $x$, which implies
\begin{align}
\la \psi|\sin(p\cdot x)\psi\ra=0.\label{VaniSin}
\end{align} 
Using the elementary fact that  $1-\cos\theta=2\Big\{\sin(\theta/2)\Big\}^2$,
we have,  by (\ref{VaniSin}), 
\begin{align}
1-(2\pi)^{d/2} \hat{\varrho}(p)=\la \psi|(\one -e^{-i p \cdot x})\psi\ra
=2 \Big\la \psi
\Big|\Big\{\sin\Big(\frac{p\cdot x}{2}\Big)\Big\}^2 \psi\Big\ra. \label{SINEEX}
\end{align} 
Note that    the multiplication operator $\Big\{\sin\Big(\frac{p\cdot
x}{2}\Big)\Big\}^2$ satisfies $
\Big\{\sin\Big(\frac{p\cdot x}{2}\Big)\Big\}^2\unrhd 0
$
w.r.t. $L^2(\BbbR^d; dx)_+$,  and is  nonzero if and only if $p\neq 0$.
Hence, by Proposition \ref{GSUni} (i) and  Theorem \ref{VecP3},  the RHS of (\ref{SINEEX})
is strictly positive if and only if $p\neq 0$.

(iii) Note that if $f\in \mathfrak{A}_{\mathrm{e}}$, then $\overline{f}\in
\mathfrak{A}_{\mathrm{e}}$ as well.
Thus, by Theorem \ref{GriSec} (i), we have
\begin{align}
\la fg\ra\ge \la f\ra\la g\ra,\ \  \la f \overline{g}\ra\ge \la f\ra\la \overline{g}\ra.
\end{align}  
Since $\la \overline{g}\ra=\la g\ra$, 
\begin{align}
\la fg\ra+\la f\overline{g}\ra\ge 2\la f\ra\la g\ra. 
\end{align} 
Let $C_0(\BbbR^d)$ be the set of all continuous
functions on $\BbbR^d$ with compact support.
Observe that, for all $f, g\in \mathfrak{A}_{\mathrm{e}} \cap
C_0(\BbbR^d)$, 
\begin{align}
\la fg\ra&=(2\pi)^{-d/2} \int_{\BbbR^d\times \BbbR^d} dpdp'
 \hat{f}(p)\hat{g}(p') \hat{\varrho}(p+p'),\\
\la f\overline{g}\ra&=(2\pi)^{-d/2} \int_{\BbbR^d\times \BbbR^d} dpdp'
 \hat{f}(p)\hat{g}(p') \hat{\varrho}(p-p')
\end{align} 
and 
\begin{align}
\la f\ra\la g\ra
=\int_{\BbbR^d\times \BbbR^d} dpdp'
 \hat{f}(p)\hat{g}(p') \hat{\varrho}(p)\hat{\varrho}(p').
\end{align} 
Since $\hat{\varrho}(p)>0,\ \hat{f}(p) \ge 0$ and $\hat{g}(p)\ge 0$
for  all $f, g\in \mathfrak{A}_{\mathrm{e}} \cap
C_0(\BbbR^d)$, we arrive at 
\begin{align}
(2\pi)^{-d/2}\{
\hat{\varrho}(p+p')
+\hat{\varrho}(p-p')
\}\ge 2\hat{\varrho}(p)\hat{\varrho}(p').
\end{align}
This completes the proof of (iii). $\Box$

\subsection{Proof of Theorems \ref{App2} and \ref{App3} }
These theorems follow immediately  from Theorems \ref{MonoEx} and
\ref{MonoG}. $\Box$

\appendix
\setcounter{equation}{0}

\section{General theory of correlation inequalities}\label{AppA}

In this appendix, we will review some basic results concerning the operator inequalities introduced 
in Section \ref{3}. Almost all results are taken from the author's previous works \cite{Miyao1,Miyao2,  Miyao4, Miyao5, Miyao6, Miyao7, Miyao8}.

\begin{Prop}\label{Basic}
Let $\{A_n\}_{n=1}^{\infty}\subseteq \mathscr{B}(\h)$  and let $A\in
 \mathscr{B}(\h)$.  Suppose that $A_n$ converges to $A$ in the weak
 operator topology. If $A_n\unrhd 0$ w.r.t. $\Cone$ for all $n\in
 \BbbN$, then $A\unrhd 0$ w.r.t. $\Cone$.
\end{Prop} 
{\it Proof.} By Remark \ref{Pequiv} (i), $\la \xi|A_n\eta\ra\ge 0$
for all $\xi, \eta\in \Cone$. Thus, 
$\displaystyle 
\la \xi|A\eta\ra=\lim_{n\to \infty}\la \xi|A_n\eta\ra\ge 0
$ for all $\xi, \eta\in \Cone$. By Remark \ref{Pequiv} (i) again, we
conclude that $A\unrhd 0$ w.r.t. $\Cone$. $\Box$
\medskip\\

\begin{Prop}\label{GSPosi}
Let $A$ be a self-adjoint positive operator on $\h$. Assume that $
e^{-\beta A} \unrhd  0$ w.r.t. $\Cone$ for all $\beta \ge 0$.
Assume that $E=\inf \sigma(A)$ is an eigenvalue of $A$.
Then there exists a nonzero vector $\xi\in \ker(A-E)$ such that $\xi\ge
 0$
w.r.t. $\Cone$.
\end{Prop} 
{\it Proof.} 
Let $\eta\in \h$. By Theorem \ref{SAH}, we can express $\eta$ as
$\eta=\eta_{R}+i \eta_{I}$ with $\eta_{R},
\eta_{I} \in \h_{\BbbR}$. Now, we define an antilinear
involution $J$ by $J\eta=\eta_{R}-i \eta_{I}$.
Clearly, 
\begin{align}
\eta_{R}=\frac{1}{2}(\eta+J\eta),\ \ \
 \eta_{I}=\frac{1}{2i}
(\eta-J\eta). \label{RI}
\end{align} 
Moreover, $\h_{\BbbR}=\{\eta\in \h\, |\, J\eta=\eta\}$. Because $e^{-\beta
A} \Cone \subseteq \Cone$, we see that $e^{-\beta A} \h_{\BbbR}
\subseteq \h_{\BbbR}$ for all $\beta \ge 0$, see  Remark \ref{Pequiv} (i).
Hence, for all $\beta \ge 0$, we obtain
\begin{align}
Je^{-\beta A} =e^{-\beta A}J. \label{JComm}
\end{align} 

Let $\xi\in \ker(A-E)$ with $\xi\neq 0$. $\xi$ can be expressed as
$\xi=\xi_{R}+i \xi_{I}$ with $\xi_{R},
\xi_{I} \in \h_{\BbbR}$. Because $\xi\neq 0$, we have
$\xi_{R} \neq 0$ or $\xi_{I} \neq 0$.
By (\ref{RI}) and (\ref{JComm}), we know that $\xi_{R},
\xi_{I}\in \ker(A-E) \cap \h_{\BbbR}$.
Without loss of generality, we may assume that $\xi_{R} \neq
0$.
By Definition \ref{HilCone} (ii)  and Theorem \ref{SAH}, we have a unique
decomposition $\xi_{R}=\xi_{R, +}-\xi_{R,
-}$,
 where $\xi_{R, \pm} \in \Cone$ with $\la \xi_{R,
 +}|\xi_{R, -}\ra=0$.
Let $|\xi_{R}|=\xi_{R, +}+\xi_{R, -}$.
Because $\|\xi_{R}\|=\||\xi_{R}|\|$,  we have
\begin{align}
e^{-\beta E} \|\xi_{R}\|^2=\la \xi_{R}|e^{-\beta A}
 \xi_{R} \ra
\le 
\la |\xi_{R}||e^{-\beta A}
| \xi_{R}| \ra
\le 
e^{-\beta E} \|\xi_{R}\|^2.
\end{align} 
Thus, $|\xi_{R}| \in \ker(A-E)$. Clearly, $|\xi_{R}|
\ge 0$ w.r.t. $\Cone$. $\Box$

\begin{Thm}\label{StandPP}
Let $A$ be a  self-adjoint positive operator on $\h$ and $B\in \mathscr{B}(\h)$. 
Suppose that 
\begin{itemize}
\item[{\rm (i)}] $e^{-\beta A} \unrhd 0$ w.r.t. $\Cone$ for all $\beta
	     \ge 0$;
\item[{\rm (ii)}] $B\unrhd 0$ w.r.t. $\Cone$.
\end{itemize} 
Then we have $e^{-\beta (A-B)}\unrhd 0$ w.r.t. $\Cone$ for all $\beta
 \ge 0$.
\end{Thm}
{\it Proof.} By (ii) and Proposition \ref{Basic}, 
\begin{align}
e^{\beta B}=\sum_{n\ge 0} \underbrace{\frac{\beta^n}{n!}}_{\ge 0} 
\underbrace{B^n}_{\unrhd 0} \unrhd 0\ \ \mbox{w.r.t. $\Cone$ for all
 $\beta \ge 0$.}
\end{align} 
Hence, by (i) and Proposition \ref{Miura} (ii),
\begin{align}
\Big(
\underbrace{e^{-\beta A/n}}_{\unrhd 0} 
\underbrace{e^{\beta B/n}}_{\unrhd 0}
\Big)^n \unrhd 0\ \ \ \mbox{w.r.t. $\Cone$ for all $\beta \ge 0$.}
\end{align}
Using the Trotter--Kato product formula(e.g.,  \cite[Theorem S. 21]{ReSi1}) and Proposition \ref{Basic}, we
arrive at the desired assertion. $\Box$

\begin{Thm}\label{Mono}
Let $A, B$ be self-adjoint positive 
 operators on $\h$. 
Assume that $B=A-C$ with $C\in \mathscr{B}(\h)$.
Suppose 
 that 
\begin{itemize}
\item[{\rm (i)}] $e^{-\beta A}\unrhd 0$ w.r.t. $\Cone$ for
	     all $\beta\ge 0$;
\item[{\rm (ii)}] $C\unrhd 0$ w.r.t. $\Cone$.
\end{itemize} 
Then we have $e^{-\beta B }\unrhd e^{-\beta A}$
 w.r.t. $\Cone$ for all $\beta\ge  0$. 
\end{Thm} 
{\it Proof.}  By the Duhamel formula, we have the  norm-convergent expansion
\begin{align}
e^{-\beta B}&=\sum_{n=0}^{\infty}D_n(\beta), \label{Duha1}\\
D_n(\beta)&=\int_{S_n(\beta)} e^{-s_1 A}C e^{-s_2 A}C\cdots
 e^{-s_n A} C e^{-(\beta-\sum_{j=1}^ns_j)A},
\end{align} 
where $\int_{S_n(\beta)}=\int_0^{\beta}ds_1\int_0^{\beta-s_1}ds_2\cdots
\int_0^{\beta-\sum_{j=1}^{n-1}s_j} ds_n$ and
$D_0(\beta)=e^{-\beta A}$. Since $C \unrhd 0$ and $\ex^{-t
A}\unrhd 0$ w.r.t. $\Cone$ for all $t \ge 0$, it holds that, by Proposition \ref{Miura} (ii),
\begin{align}
\underbrace{ 
e^{-s_1 A}
}_{\unrhd 0}
\underbrace{C}_{\unrhd 0}
\underbrace{ 
e^{-s_2 A}
}_{\unrhd 0}\cdots
 \underbrace{e^{-s_n A}}_{\unrhd 0} 
\underbrace{C}_{\unrhd 0}
\underbrace{e^{-(\beta-\sum_{j=1}^ns_j)A}}_{\unrhd 0} \unrhd 0
\end{align} 
provided that $s_1 \ge 0, \dots, s_n\ge 0$ and $\beta-s_1-\cdots-s_n\ge 0$. Thus, by Proposition \ref{Basic},
we obtain
$D_n(\beta)\unrhd 0$ w.r.t. $\Cone$ for all $n\ge 0$.
 Accordingly,  by (\ref{Duha1}) and Proposition \ref{Basic} again, we have $\ex^{-\beta B}\unrhd
 D_{n=0}(\beta)=e^{-\beta A}$ w.r.t. $\Cone$ for all
 $\beta \ge 0$. $\Box$
\medskip

\begin{rem}
{\rm 
By (i), there exists a unique $\xi\in \h$ such that $\xi>0$
 w.r.t. $\Cone$
and $P_A=|\xi\ra \la \xi|$. 
Of course, $\xi$ satisfies $A\xi=\inf \sigma(A)\xi$.
$\diamondsuit$
}
\end{rem}

\begin{Thm}\label{ErgPI}
Let $A$ be a self-adjoint positive operator on $\h$, and let
 $B\in \mathscr{B}(\h)$.  
Suppose the following:
\begin{itemize}
\item[{\rm (i)}] $e^{-\beta A} \unrhd 0$ w.r.t. $\Cone$ for all $\beta \ge 0$.
\item[{\rm (ii)}] $B$ is ergodic w.r.t. $\Cone$.
\end{itemize}
Then,  $e^{-\beta (A-B)} \rhd 0$ w.r.t. $\Cone$ for all $\beta>0$.
\end{Thm}
{\it Proof.} 
Set $H=A-B$.
We apply  Fr\"ohlich's idea \cite{JFroehlich1} and use the Duhamel expansion:
\begin{align}
e^{-\beta H}&=\sum_{n\ge 0} \mathscr{D}_n(\beta), \label{D0}\\
\mathscr{D}_n(\beta) &= \int_{S_n(\beta)} e^{-s_1 A} B e^{-s_2 A} \cdots e^{-s_n A} B
e^{-(\beta-\sum_{j=1}^n s_j) A}.
\end{align}
In a manner  similar to that used in   the proof of Theorem \ref{Mono}, we know that 
\begin{align}
&\mathscr{D}_n(\beta) \unrhd 0, \label{D1}\\
&e^{-s_1 A} B e^{-s_2 A} \cdots e^{-s_n A} B
e^{-(\beta-\sum_{j=1}^ns_j) A} \unrhd 0 \label{D2}
\end{align}
w.r.t. $\Cone$, provided that $s_1 \ge 0, \dots, s_n\ge 0$ and $\beta-s_1-\cdots-s_n\ge 0$.

Let $\xi, \eta\in \Cone\backslash \{0\}$. Since $e^{-\beta A} \unrhd 0$ w.r.t. $\Cone$
 for all $\beta \ge 0$, we have $e^{-\beta A} \eta\in \Cone\backslash \{0\}$.
Let $\beta>0$ be fixed arbitrarily.
Because $B$ is ergodic w.r.t. $\Cone$, there exists an $n\in \{0\}\cup \BbbN$
such that $
\la \xi|B^n\, e^{-\beta A} \eta\ra>0
$.
Now,  let
\begin{align}
F(s_1, \dots, s_n)=\Big\la \xi \Big|e^{-s_1 A} B e^{-s_2 A} \cdots e^{-s_n A} B
e^{-(\beta-\sum_{j=1}^ns_j) A} \eta\Big\ra.
\end{align}
By (\ref{D2}), it holds that $F(s_1, \dots, s_n) \ge 0$.
In addition, we have $
F(0, \dots, 0)=\la \xi|B^n e^{-\beta A}\eta\ra>0
$.
Because $F(s_1,\dots, s_n)$ is continuous in $s_1, \dots, s_n$, we obtain
\begin{align}
\la \xi|\mathscr{D}_n(\beta) \eta\ra
=\int_{S_n(\beta)} F(s_1, \dots, s_n) >0.
\end{align}
By (\ref{D0}) and (\ref{D1}), we see that $
e^{-\beta H} \unrhd \mathscr{D}_n(\beta)
$, which implies 
\begin{align}
 \la \xi|e^{-\beta H}\eta\ra \ge \la \xi|\mathscr{D}_n(\beta)\eta\ra>0.
\end{align}
Since $\xi$ and $\eta$ are in $\Cone\backslash\{0\}$, we conclude that $e^{-\beta H}
\eta>0$ w.r.t. $\Cone$.
 Since $\beta$ is arbitrary, we obtain that $e^{-\beta H} \rhd 0$ w.r.t. $\Cone$
 for all $\beta>0$.  $\Box$

\begin{Thm}\label{VecP3}
Let $A\in \mathscr{B}(\h)$. Assume that $u>0$ w.r.t. $\Cone$ and  
$A\unrhd 0$ w.r.t. $\Cone$.
Then,  $\la u|Au\ra=0$ if and only if $A=0$.
\end{Thm} 
{\it Proof.} We will divide the proof into several steps.

\begin{flushleft}
{\bf Step 1.}
{\it Let $A\in \mathscr{B}(\h)$. If $Au= 0$ for all $u\in \Cone$, then $A=0$.}
\end{flushleft}
{\it Proof.} By Remark \ref{DecRI}, each $u\in \h$ can be written as 
$
u=v_1-v_2+i(w_1-w_2)
$, where $v_1, v_2, w_1, w_2\in \Cone$ such that $\la v_1|v_2\ra=0$ and
$\la w_1|w_2\ra=0$. Thus, the assumption implies that $Au=0$
 for {\it all} $u\in \h$. $\Box$

\begin{flushleft}
{\bf Step 2.} {\it 
Let $A\in \mathscr{B}(\h)$ with $A\neq 0$. Assume that  $u>0$ w.r.t. $\Cone$.
If $A\unrhd 0$ w.r.t. $\Cone$, then $Au\neq 0$.}
\end{flushleft} 
{\it Proof.}
Assume that $Au=0$. Then,  $\la v|Au\ra=0$ for all $v\in \Cone$,
implying that  $\la A^*v|u\ra=0$.
Since $u>0$  and $A^* v\ge 0$ w.r.t. $\Cone$, we conclude that 
$A^*v$ must be zero. Because $v$ is arbitrary, 
$A^*=0$  by {\bf Step 1}. $\Box$
\begin{flushleft}
{\it Completion of the proof.}
\end{flushleft} 
Suppose that $\la u|Au\ra=0$. Assume that $A\neq 0$.
Since $Au \ge 0$ and $u>0$ w.r.t. $\Cone$, $Au$ must be zero.
 However,  this contradicts with   {\bf Step 2}. $\Box$

\end{document}